\documentclass[journal,letter,compsoc]{IEEEtran}						

\usepackage{cite}
\usepackage[cmex10]{amsmath} 

\usepackage{amssymb,dblfloatfix}

\usepackage[hidelinks]{hyperref}

\usepackage{graphicx}
\usepackage{dcolumn}
\usepackage{bm}

\usepackage{algorithm}
\usepackage{color}
\usepackage{tikz}	
\usetikzlibrary{backgrounds,fit,arrows,decorations.pathreplacing,positioning}

\usepackage{enumitem}	

\newtheorem{definitionenv}{Definition}
\newtheorem{lemmaenv}[definitionenv]{Lemma}
\newtheorem{theoremenv}[definitionenv]{Theorem}
\newtheorem{corollaryenv}[definitionenv]{Corollary}
\newtheorem{propositionenv}[definitionenv]{Proposition}
\newtheorem{remarkenv}[definitionenv]{Remark}
\newtheorem{conjectureenv}[definitionenv]{Conjecture}
\newtheorem{exampleenv}{Example}
\newtheorem{app-lemmaenv}[section]{Lemma}
\newenvironment{definition}{\begin{definitionenv}\rm}{\end{definitionenv}}
\newenvironment{lemma}{\begin{lemmaenv}\rm}{\end{lemmaenv}}

\newenvironment{remark}{\begin{remarkenv}\rm}{\end{remarkenv}}

\newenvironment{app-lemma}{\begin{app-lemmaenv}\rm}{\end{app-lemmaenv}}

\DeclareMathOperator*{\argmax}{arg\,max}
\DeclareMathOperator*{\argmin}{arg\,min}
\DeclareMathOperator{\sgn}{sign}

\newcommand{\sM}{{\cal M}}
\newcommand{\sN}{{\cal N}}
\newcommand{\sS}{{\cal S}}

\newcommand{\cM}{{\cal M}}
\newcommand{\cN}{{\cal N}}
\newcommand{\cS}{{\cal S}}

\newcommand{\mR}{{\mathbb R}}

\usetikzlibrary{backgrounds,fit,arrows,decorations.pathreplacing,positioning}
\usepackage[caption=false,labelformat=simple]{subfig}		

\usepackage[colorinlistoftodos,prependcaption]{todonotes}
\usepackage{xcolor}
\usepackage{xargs}

\newcommandx{\yellownote}[2][1=]{\todo[inline,linecolor=yellow,backgroundcolor=yellow!25,bordercolor=yellow,#1]{#2}}

\newcommand{\ourBP}{{MBP$_4$}\xspace}	\usepackage{xspace}

\usepackage[normalem]{ulem}

\newcommand{\eq}[1]{\eqref{#1}}		

\begin{document}


\title{Exploiting degeneracy in belief propagation decoding of quantum codes}

%
%
%

\author{Kao-Yueh~Kuo 
    and Ching-Yi~Lai
\thanks{K.-Y.~Kuo and C.-Y.~Lai are with the Institute of Communications Engineering, National Yang Ming Chiao Tung University, Hsinchu 30010, Taiwan. 
	(e-mail: kykuo@nycu.edu.tw and cylai@nycu.edu.tw)}
}

\maketitle

\begin{abstract}
Quantum information needs to be protected by quantum error-correcting codes due to imperfect physical devices and operations. One would like to have an efficient and high-performance decoding procedure for the class of quantum stabilizer codes. A potential candidate is Pearl's belief propagation (BP), but its performance suffers from the many short cycles inherent in a quantum stabilizer code, especially highly-degenerate codes. A general impression exists that BP is not effective for topological codes. In this paper, we propose a decoding algorithm for quantum codes based on quaternary BP  with additional memory effects (called MBP). This MBP is like a recursive neural network with inhibitions between neurons (edges with negative weights), which enhance the perception capability of  a network. Moreover, MBP exploits the degeneracy of a quantum code so that the most probable error or its degenerate errors can be found with high probability. The decoding performance is significantly improved over the conventional BP for various quantum codes, including quantum bicycle, hypergraph-product,  surface and toric codes. For MBP on the surface and toric codes over depolarizing errors, we observe error thresholds of 16\% and 17.5\%, respectively.
\end{abstract}
\begin{IEEEkeywords}
belief propagation, degeneracy, sparse-graph codes, quantum stabilizer codes, surface and toric codes, error threshold.
\end{IEEEkeywords}


%


\section{Introduction} \label{sec:Intro}

To demonstrate an interesting quantum algorithm, such as Shor's factoring algorithm \cite{Shor94}, a quantum computer needs to implement more than $10^{10}$ logical operations, which means that the error rate of each logical operation must be much less than $10^{-10}$ \cite{SKFCLP13}.
With limited quantum devices and imperfect operations \cite{WUZ+17,AAB19}, quantum information needs to be protected by quantum error-correcting codes to achieve fault-tolerant quantum computation \cite{Shor96}.
If a quantum state is encoded in a stabilizer code \cite{GotPhD,CRSS98}, the \emph{error syndrome} of an occurred error can be measured without disturbing the quantum information of the state.  
A quantum stabilizer code constructed from a sparse graph is favorable since it affords a two-dimensional layout or simple quantum error-correction  procedures.
This includes the families of  surface and toric codes \cite{Kit03}, color codes \cite{BM06},
bicycle codes \cite{MMM04}, and  generalized hypergraph-product (GHP) codes \cite{TZ14,KP13}.

For a general stabilizer code, the decoding problem of finding the most probable coset of degenerate errors with a given error syndrome is  hard \cite{KL13_20,IP15},
and an efficient decoding procedure with good performance is desired.
The complexity of a decoding algorithm is usually a function of code length $N$.
Edmonds' minimum-weight perfect matching (MWPM) \cite{Edm65} can be used to decode a surface or toric code \cite{DKLP02,WHP03,RHG06,WFSH10}.
The complexity of MWPM is $O(N^3)$, and can be reduced  to $O(N^2)$ if local matching is used with minor performance loss \cite{WFSH10,FWH12,Fow15}.
Duclos-Cianci and Poulin proposed a renormalization group (RG) decoder, which uses a strategy analogous to the decoding of a concatenated code, to decode a toric (or surface) code with complexity proportional to $N\log(\sqrt{N})$ \cite{DP10}.
Both MWPM and RG can be generalized for color codes (see Appendix~\ref{sec:color}).

On the other hand, most sparse-graph quantum codes can be decoded by belief propagation (BP) \cite{MMM04,KL20,KL20b,KL21a} or its variants.\footnote{
	See: BP with random perturbation (tested on a bicycle code) \cite{PC08}, 
	RG-BP on toric codes \cite{DP10}, RG-BP on color codes \cite{BDP12,SR12}, 
	BP-MWPM on surface/toric codes \cite{CA18}, BP-OSD on (generalized) bicycle, HP and topological codes \cite{PK19,RWBC20},
	and BP with small set flipping (BP-SSF) on HP codes with expander graphs \cite{GGKL21} (where SSF \cite{LTZ15,FGL18a} is a generalization of {\it bit-flipping BP} for classical expander codes \cite{SS96}).
	}
BP is an iterative algorithm with complexity $O(Nj)$ per iteration \cite{DM98b,KL21a}, 
where $j$ is the mean column-weight of the check matrix of a quantum code. 
In general, an average number of iterations $\tau\approx \log\log N$ is sufficient for BP to converge \cite{Gal63,Mac99}.
In practice, a maximum number of iterations $T_{\max}$ proportional to $\tau$ up to a large enough constant will be chosen.
So the complexity of BP is $O(NjT_{\max})$ or more precisely $O(Nj\tau) = O(Nj\log\log N)$.

Although BP seems to have the lowest complexity, a long-standing problem  
is that  BP does not perform well on quantum codes with {high degeneracy},
unless additional complex processes are included \cite{CA18,PK19,RWBC20}.
(We say that a code has \emph{high degeneracy} or is \emph{highly-degenerate} if it has many  stabilizers of weight lower than its minimum distance.) 
The Tanner graph of a stabilizer code inevitably contains many short cycles, which deteriorate the message-passing process in BP \cite{MMM04,PC08}, 
especially for codes with high degeneracy \cite{PK19,RWBC20,RV20}.
Any message-passing or neural network decoder may suffer from this issue.
One may consider variants of BP with additional efforts in pre-training by neural networks \cite{TM17,KJ17,LP19,MKJ19} or post-processing \cite{PK19,RWBC20} such as ordered statistics decoding (OSD) \cite{FL95}, but these methods may not be practical for large codes.
In this paper we will address this long-standing BP problem by devising an efficient quaternary BP 
decoding algorithm with additional memory effects (cf.~\eq{eq:gamma_nm1}), abbreviated \emph{MBP}, so that the degeneracy of quantum codes can be exploited.

Many known decoders in the literature  treat Pauli $X$ and $Z$ errors separately as binary errors, 
which may incur additional computation overhead or performance loss. 
MBP directly handles the quaternary errors.

The problem of a hard-decision decoding of a classical code is like an energy-minimization problem in a neural network \cite{BB89},
where an energy function measures the parity-check satisfaction (denoted by $J_\text{S}$).
It is known that
BP has been used for energy minimization in statistical physics \cite{Mac99,PC08,YFW05}.
Moreover, an iterative decoder based on the gradient decent optimization of the energy function has been proposed \cite{LBB98}. 
These motivate us to consider a soft-decision generalization of the energy function 
with variables that are log-likelihood ratios (LLRs) of Pauli errors
and make connections between BP and   the gradient decent algorithm.
We define an energy function with an additional term (denoted by $J_\text{D}$) that measures the distance between a recovery operator and the initial channel statistics.
Then we show that BP in log domain is like a gradient descent optimization for this generalized energy function (cf.~\eq{eq:energy}) 
but with more elegant step updates. 
This explains why  the conventional BP usually works on a nondegenerate quantum code, 
since the energy topology is similar to the classical case  
(cf.~Sec.~\ref{sec:J_S}).

For a highly-degenerate quantum code, it has many low-weight stabilizers corresponding to local minimums in the energy topology 
so that the conventional BP easily gets trapped in these local minimums near the origin (see~Fig.~\ref{fig:BDD}).
This suggests that we should use a larger step (which can be controlled by message normalization) \cite{KL21a}.
However, this is simply not enough since the energy minimization process may not converge if large steps are made.
An observation from Hopfield nets is that \emph{inhibitions} (edges with negative weights) between neurons can enhance the perception capability of a network and improve the pattern-recognition accuracy \cite{Hop84,HT85,HT86,BM89,MWW91}.
MBP is mathematically formulated to have this inhibition functionality, which helps to resist wrong beliefs passing on the Tanner graph (due to short cycles \cite{Gal63}) or to effectively accelerate the search in a gradient descent optimization to avoid getting trapped.
An important feature of MBP is that no additional computation is required and thus the complexity of MBP remains the same as the conventional BP with  message normalization.

The performance of MBP can be further improved by choosing an appropriate step-size for each error syndrome. However, it is difficult to precisely determine the step-size. If the step-size is too large, MBP may return incorrect solutions or diverge.  
We propose to adaptively choose the step-size using an $\varepsilon$-net. 
This adaptive scheme will be called \emph{AMBP}, and its complexity is still $O(Nj\log\log N)$ since the chosen $\varepsilon$ is independent of $N$.

An optional technique that can adopted in MBP is to use \emph{fixed initialization} \cite{HFI12,KL21a}.
The energy function and energy topology are defined according to the channel statistics. 
If MBP performs well on a certain channel statistics (say, at a certain depolarizing rate $\epsilon_0$), 
it means that MBP can correctly determine most syndrome-and-error pairs on that topology.
Thus it is better to decode using this energy topology, regardless of the true channel statistics.  
This technique works for any quantum codes.

Computer simulations of MBP (or AMBP) on various quantum codes are performed. 
Note that MBP naturally extends to a model of simultaneous data and measurement errors \cite{KCL21};
however, perfect syndrome measurements are assumed in this paper. 
We will first do a case study on the five-qubit code \cite{LMPZ96} to show how the memory effects help the gradient descent optimization. 
Then we decode quantum bicycle codes \cite{MMM04}, a GHP code in \cite{PK19}, and (rotated) surface and toric codes \cite{BM07,HFDM12}.
	
	Bicycle codes have good error-correction performance and low decoding complexity but they may have high error-floor if there are many low-weight stabilizers \cite[Fig.~6]{MMM04}, which occurs if the generator vector of a bicycle code has low weight. 
	This vector generates a bicycle check matrix with fixed stabilizer weight, called as \textit{row-weight}. 
	A bicycle code has its minimum distance upper bounded by its row-weight.
	However, the performance of BP depends more on the weight distribution of codewords \cite{Gal63,Mac99}, rather than just the code distance.
	We simulate the cases in \cite[Fig.~6]{MMM04} and each error-floor is significantly improved using MBP. 
	In particular, the convergence behavior can be improved, especially when the row-weight is small (i.e., when the code is more degenerate).
	If AMBP is used, the achieved performance is close to the quantum Gilbert--Varshamov rate (see Fig.~\ref{fig:AMBP_bic} and its discussions.)

 Next, we consider a  GHP code constructed in \cite{PK19} with parameters $[[N,K,D]]=[[822,48,16]]$ and row-weight~8.
	Since the row-weight $8<D$, 
	the code is highly-degenerate so the energy topology is hard for conventional BP to have good convergence.
	In \cite{PK19}, a high order \textit{BP-OSD-$\omega$} with order $w$ is proposed. 
	BP is used with post-processing by OSD, together with a post-selection on $2^\omega$ possible errors.
	The considered GHP code needs BP-OSD-$w$ with $w=15$ so that degenerate errors can be found with high probability.
	OSD has subroutines of sorting, Gaussian elimination, and classical re-encoding \cite{FL95}. 
	Thus, further with the post-selection on $2^w$ errors, the complexity of BP-OSD-$\omega$ is quite high.
	On the other hand, AMBP efficiently finds degenerate errors and can outperform BP-OSD-15 (see Fig.~\ref{fig:GHP}).

Finally we consider the decoding \textit{threshold} on surface codes  ($\epsilon_\text{surf}$) or toric codes  ($\epsilon_\text{toric}$) as a \mbox{benchmark~\cite{DKLP02}}.
Theoretical estimation suggests that surface or toric codes have a threshold of 18.9\% over depolarizing errors \cite{WL12,BAOKM12,Ohz12}, 
which agrees with the hashing bound of increasing overhead \cite{BDSW96}.
In the following, we compare decoders with practical complexity. 
MWPM achieves a threshold of 15.5\% \cite{WFSH10,WHP03}.
RG~combined with BP (RG-BP) achieves $\epsilon_\text{toric}=16.4\%$ \cite{DP10}.
A decoder based on matrix product states (MPS) achieves $17\% \le \epsilon_\text{surf} \le 18.5\%$ \cite{BSV14} 
(which has complexity $O(N^2)$ like MWPM but is more complex in practice since MPS needs many matrix operations).
Union-find (UF) has complexity almost linear in $N$ and a threshold of 9.9\% on toric codes over bit-flip errors \cite{DN17} (which after rescaled by $\frac{3}{2}$ is 14.85\% over depolarizing errors). 
BP-assisted MWPM (BP-MWPM) has a threshold of 17.76\% with complexity $O(N^{2.5})$ \cite{CA18}.
AMBP roughly achieves $\epsilon_\text{surf}=16\%$ and $\epsilon_\text{toric}=17.5\%$.
The surface and toric codes have mean column-weight $j\le 4$ \cite{Kit03,HFDM12}. 
Hence the complexity of (A)MBP is $O(Nj\tau)=O(N\log\log N)$, since $\tau=O(\log\log N)$ is good enough in our in simulations, 
which agrees with the classical expectation mentioned earlier, although we need to consider short cycles and degeneracy.
The thresholds and complexities of  various decoders are provided in Table~\ref{tbl:thrd}. 
%

\begin{table}
	\caption{ 			
		The thresholds and computation complexities of various decoders on surface codes ($\epsilon_\text{surf}$) and toric codes ($\epsilon_\text{toric}$) over depolarizing errors.
		An~entry is denoted $--$ if the data is not provided in the literature.
	} 					
	\label{tbl:thrd} \centering
	$\begin{array}{|l|l|l|l|}
	\hline
	\text{decoder}				& \epsilon_\text{surf}		& \epsilon_\text{toric}			& \text{complexity}	\\
	\hline                                              
	\text{MWPM \cite{Edm65}}	& 15.5\% \text{\cite{WFSH10}}	& 15.5\%^\text{a,b} \text{\cite{WHP03}}		& O(N^2) \text{\cite{FWH12}}	\\
	\text{RG-BP \cite{DP10}}	& --						& 16.4\%						& O(N\log N)		\\
	\text{MPS \cite{BSV14}}		& 17\%\text{--}18.5\%		& --							& O(N^2)			\\
	\text{UF \cite{DN17}}		& --						& 14.85\%^\text{a}				& O(N) \\ 
	\text{BP-MWPM \cite{CA18}}	& 17.76\%^\text{b}			& 17.76\%						& O(N^{2.5})		\\
	\text{AMBP {(this paper)}}	& 16\%						& 17.5\%						& O(N\log\log N)	\\
	\hline
	\end{array}$
	\begin{flushleft} 
	\begin{itemize}[leftmargin=*]
	\item[a.] If only the threshold of a decoder over bit-flip errors is provided, we rescale it by a factor $\frac{3}{2}$ like \cite[Eq.~(40)]{MMM04}.
	\item[b.] It is usually considered that $\epsilon_\text{surf} \le \epsilon_\text{toric}$, since a toric code does not have boundary qubits.
		Figure~10 in \cite{WFSH10} seems to suggest $\epsilon_\text{toric}=15\% < \epsilon_\text{surf}=15.5\%$ according to the intersection points.
		However, the MWPM threshold on toric codes over bit-flip errors is estimated to be 0.1031 \cite{WHP03}, which, after rescaled by $\frac{3}{2}$, 
		matches the $\epsilon_\text{toric}=15.5\%$ finally claimed in \cite{WFSH10}. 
		Figures~10 and 12 in \cite{CA18} seem to suggest $\epsilon_\text{toric}=17.76\% < \epsilon_\text{surf}=17.84\%$, which may cause confusion,  
		so a single threshold value 17.76\% was concluded in \cite{CA18}. 
	\end{itemize}
	\end{flushleft}
\end{table}

	Our simulation results show that MBP significantly improves the decoding performance of conventional BP. In particular, degeneracy is exploited so that degenerate errors may be returned by the decoder.
 It is known that BP can be treated as a recurrent neural network (RNN) \cite{Nac+18,LP19}.
 Similarly, our MBP induces an RNN with inhibition without the pre-training process.
 This may provide an explanation why RNN decoders (which may contain many negative-weight edges after training) work well on degenerate codes.

This paper is organized as follows.
In Sec.~\ref{sec:pre}, we introduce stabilizer codes and the decoding problem.
In Sec.~\ref{sec:bp_me}, we interpret the decoding problem as energy minimization and introduce our MBP algorithm.
We also show the RNN induced by MBP and link it with the inhibition technique in Hopfield nets.
In Sec.~\ref{sec:Sim}, we provide the 
simulation results of MBP decoding on various quantum codes.
Finally we conclude and discuss some future research topics  in Sec.~\ref{sec:Conclu}.

\section{Preliminaries} \label{sec:pre}

\subsection{Stabilizer codes} 

We consider errors that are tensor product of Pauli matrices 
$\left\{
I=\left[\begin{smallmatrix}1 &0\\0&1\end{smallmatrix}\right],
X=\left[\begin{smallmatrix}0 &1\\1&0\end{smallmatrix}\right],
Y=\left[\begin{smallmatrix}0 &-i\\i&0\end{smallmatrix}\right],
Z=\left[\begin{smallmatrix}1 &0\\0&-1\end{smallmatrix}\right] 
\right\}$. 
In particular, we will simulate independent depolarizing errors with rate $\epsilon\in(0,3/4)$  so that each qubit independently suffers a Pauli error 
$I,X,Y,$ or $Z$, according to a distribution
	\begin{equation} \label{eq:init_p}
	(p^I,\, p^X,\, p^Y,\, p^Z)=(1-\epsilon,\, \epsilon/3,\, \epsilon/3,\, \epsilon/3).
	\end{equation} 
The \emph{weight} of an $N$-fold Pauli operator in $\{I,X,Y,Z\}^{\otimes N}$, regardless of the global phase, is the number of its non-identity components. 
For small $\epsilon$, Pauli errors of lower weight occur with higher probability  and we would like to mitigate their effects.
In the following, the notation of tensor product $\otimes$ may be omitted if no confusion arises in our discussions.

A \textit{stabilizer group} $\sS$ is an Abelian subgroup of $\{1,-1\}\times\{I,X,Y,Z\}^{N}$ such that   $-I^{N}\notin \sS$.
Assume that $\sS$ has a set of $N-K$ independent generators $\{S_m\}_{m=1}^{N-K}$. 
For simplicity, consider $S_m\in\{I,X,Y,Z\}^N$, though a generator may still have negative phase. 
A binary $[[N,K,D]]$ \textit{stabilizer code} defined by $\cS$ is the $2^K$-dimensional subspace in $\mathbb{C}^{2^N}$ that is the joint $(+1)$-eigenspace of $\sS$ \cite{GotPhD,CRSS98}.  The parameter $D$ is called  the \textit{minimum distance} of the code and will be defined below.
The elements in $\sS$ are called \emph{stabilizers}.
Two Pauli operators either commute or anticommute with each other.
If a Pauli error anticommutes with certain stabilizers, measuring those stabilizers will return eigenvalues $-1$.
Consider the measurement result $\pm1$ to be mapped by $+1\mapsto 0$ and $-1\mapsto 1$. 
Hence the measured eigenvalues of $\{S_m\}_{m=1}^{N-K}$, after mapping, are called the \textit{error syndrome} 
of the error and a nonzero error syndrome suggests a detected error.
Since a stabilizer  has no effect on the code space, an error that is a stabilizer is harmless.
	Thus the minimum distance of the stabilizer code defined by $\cS$ is the minimum weight of a Pauli error that is harmful but cannot be detected.
	Let 
	\begin{equation} \label{eq:N(S)}
	N(\cS)=\{ E\in  \{I,X,Y,Z\}^N:\ EF=FE \ \ \forall\ F\in\cS  \}. 
	\end{equation}
	Then $D=\min\{\text{weight of } {F}: F\in N(\cS)\setminus \pm\cS\}$. 

For $F\in\cS$ and an $N$-fold Pauli error $E$, the Pauli error $EF$ and $E$ are equivalent on the code space and hence $EF$ is called a \textit{degenerate error} of $E$.
A quantum code is said to be \textit{degenerate} if there are stabilizers of weight less than its minimum distance;
otherwise, the code is \textit{nondegenerate}.  
An $[[N,K,D]]$ stabilizer code has $N-K$ independent stabilizer generators. 
We judge how degenerate a code is by the percentage of its maximum number of independent stabilizer generators of weight less than $D$.
Roughly speaking, a code is \emph{highly-degenerate} or with \emph{high degeneracy} if this percentage is high. 
For example, a surface or toric code of $D\geq 5$ is highly-degenerate since it is straightforward to find a set of $N-K$ independent generators of weight $\le 4$ \cite{Kit03,HFDM12}.

For better decoding performance of BP, $M\ge N-K$ stabilizers $\{S_m\}_{m=1}^M$ will be measured.\footnote{
	Additional stabilizers $(M > N-K)$ may provide stronger protection, e.g.,
	a toric code has every qubit equally protected by four stabilizers due to additional stabilizers.
	In the case that the syndrome measurement operations are faulty, we may need to measure additional stabilizers to obtain reliable syndrome information \cite{ALB20,KCL21}.
	}
We write $S_m = S_{m1}S_{m2}\cdots S_{mN} \in\{I,X,Y,Z\}^N$ for $m=1,2,\dots,M$,
and define an $M\times N$ \emph{check matrix} 
	$$S=[S_{mn}]\in\{I,X,Y,Z\}^{M\times N}.$$ 
The \emph{row-weight} of the $m$-th row of $S$ is referred to as the weight of the stabilizer $S_m$.
 For an error $E=E_1 E_2 \cdots E_N\in\{I,X,Y,Z\}^N$, define its \emph{binary} syndrome vector $z=(z_1 z_2 \cdots z_M)\in\{0,1\}^M$ by
 \begin{equation} \label{eq:z_m}
 z_m = \sum_{n=1}^N \langle E_n, S_{mn}\rangle \mod 2, 
 \end{equation}
 where the bilinear form $\langle F_1, F_2 \rangle=0$,  if two Pauli operators $F_1$ and $F_2$ commute, and $\langle F_1, F_2 \rangle=1$, otherwise. 
 
  \begin{definition} \label{def:BDD}
  	For a code of length $N$ and minimum distance $D$, let $r\times$BDD denote the bounded distance decoding (BDD) of the code so that 
  	any Pauli errors of weight no larger than $t=\lfloor\frac{rD-1}{2}\rfloor$ are correctable. 
  	We say that $r\times$BDD has correction radius $t$ and decoding error rate
		$$ P_\text{BDD}(t) = 1 - \textstyle \left( \sum_{j=0}^t  \binom{N}{j} \epsilon^{j}(1-\epsilon)^{N-j} \right) $$
	at depolarizing error rate $\epsilon$.
  \end{definition}
	We use BDD to denote $1\times$BDD for simplicity.
Usually a good decoding procedure on a classical code has performance between BDD and $2\times$BDD. 
However, the degeneracy of a quantum code is not considered in BDD;
we may have decoding performance much better than $2\times$BDD in the quantum case.
In addition, since we do not know how to estimate the exact channel fidelity for large quantum codes, $r\times$BDD serves as a good benchmark.

 \subsection{BP decoding of quantum codes} \label{sec:BP}

Codes based on low-density parity-check (LDPC) matrices (also known as sparse-graph codes) achieves near channel-capacity performance in classical coding theory \cite{Gal63,Mac99}.
The parity-check matrix of a code can be depicted as a Tanner graph, which is a bipartite graph containing variable nodes and check nodes, connected properly by edges \cite{Tan81,Wib96}. 
Belief propagation (BP) on the Tanner graph iteratively passes messages between the nodes so that an estimate of the marginal distribution of the error at each coordinate can be approximated \cite{Pea88,KFL01}.

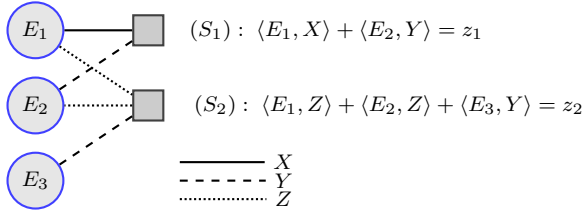
\begin{figure}[h]
	\centering~~~~ \begin{tikzpicture}[node distance=1.3cm,>=stealth',bend angle=45,auto]

\tikzstyle{chk}=[rectangle,thick,draw=black!75,fill=black!20,minimum size=4mm]
\tikzstyle{var}=[circle,thick,draw=blue!75,fill=gray!20,minimum size=4mm,font=\footnotesize]
\tikzstyle{VAR}=[circle,thick,draw=blue!75,fill=blue!20,minimum size=5mm,font=\footnotesize]
\tikzstyle{fac}=[anchor=west,font=\footnotesize]

\node[var] (x3) at (0,0) {$E_3$};
\node[var] (x2) at (0,1) {$E_2$};
\node[var] (x1) at (0,2) {$E_1$};
\node[chk] (c1) at (1.5,2) {};
\node[chk] (c2) at (1.5,1) {};

\draw[thick] (x1) -- (c1);
\draw[thick,dashed] (x2) -- (c1);
\draw[thick,densely dotted] (x1) -- (c2) -- (x2);
\draw[thick,dashed] (x3) -- (c2);


\node[fac] [right of=c1,xshift=12mm] {$(S_1):~\langle E_1,X\rangle+\langle E_2,Y\rangle = z_1$};
\node[fac] [right of=c2,xshift=19mm] {$(S_2):~\langle E_1,Z\rangle+\langle E_2,Z\rangle+\langle E_3,Y\rangle = z_2$};

\node[fac] (Xl) [right of=x3,xshift=5mm,yshift=7] {};
\node[fac] (Xr) [right of=x3,xshift=20mm,yshift=7] {$X$};
\draw[thick] (Xl) -- (Xr);
\node[fac] (Yl) [right of=x3,xshift=5mm,yshift=0] {};
\node[fac] (Yr) [right of=x3,xshift=20mm,yshift=0] {$Y$};
\draw[thick,dashed] (Yl) -- (Yr);
\node[fac] (Zl) [right of=x3,xshift=5mm,yshift=-7] {};
\node[fac] (Zr) [right of=x3,xshift=20mm,yshift=-7] {$Z$};
\draw[thick,densely dotted] (Zl) -- (Zr);

\end{tikzpicture}
	\caption[Roots associated to the Cartan matrix]{	
 		The Tanner graph 
 		induced by   $S=\left[\begin{smallmatrix} X&Y&I\\ Z&Z&Y \end{smallmatrix}\right]$. 
 		There are three variable nodes (represented by circles), two check nodes (represented by squares), and five edges. 
		Every edge has a type $X$, $Y$, or $Z$.
 	} \label{fig:S2x3}
 \end{figure}

  Consider an $M\times N$ check matrix $S\in\{I,X,Y,Z\}^{M\times N}$. The relations between a Pauli error operator and its syndrome bits  can be depicted by a  {Tanner graph} with $N$ variable nodes (corresponding to the $N$-fold Pauli error to be estimated) and $M$ check nodes (corresponding to the given binary syndrome vector) such that an edge $(m,n)$ of type $S_{mn}$
  connects variable node $n$ and check node $m$ whenever $I\ne S_{mn}\in\{X,Y,Z\}$ \cite{PC08,KL20}.  
 Figure~\ref{fig:S2x3} illustrates   the Tanner graph for a check matrix $S = \left[\begin{smallmatrix} X&Y&I\\ Z&Z&Y \end{smallmatrix}\right]$.

 Given a syndrome vector $z\in\{0,1\}^M$, the decoding problem is to find the most probable Pauli error, or one of its degenerate errors, in $\{I,X,Y,Z\}^N$.
 A quaternary BP (BP$_4$) algorithm computes an approximated marginal distribution $\hat P(E_n=W\mid z)=q_n^W$ for $W\in\{I,X,Y,Z\}$ for $n=1,2,\dots,N$ 
 in linear domain \cite{KL20}   and outputs 
	$\hat{E}=(\hat E_1, \hat E_2, \dots, \hat E_N)$ such that $$\hat E_n = \argmax_{W\in\{I,X,Y,Z\}} \hat P(E_n=W\mid z).$$
 The computation can also be done in log domain \cite{KL21a} by using log-likelihood ratios (LLRs) defined by
	\begin{align}
	\Gamma_n^X=\ln \frac{q_n^I}{q_n^X},\quad  \Gamma_n^Y=\ln \frac{q_n^I}{q_n^Y},\quad   \Gamma_n^Z=\ln \frac{q_n^I}{q_n^Z},  \label{eq:LLRs}
	\end{align}
for $n=1,2,\dots,N$.
 If the syndrome of  $\hat{E}$ matches $z$, then the decoder will output $\hat{E}$.
Since $z$ is binary, estimating $\hat E$ can be efficiently calculated  by  passing \textit{scalar messages} on the Tanner graph in linear domain \cite{KL20,KL20b} or log domain \cite{KL21a}.

In this paper we will discuss BP$_4$ in log domain with an LLR vector  
$\Gamma= (\Gamma_1,\Gamma_2,\dots, \Gamma_N)\in \mR^{3N}$, where 
$\Gamma_{n}=(\Gamma_{n}^X,\Gamma_{n}^Y,\Gamma_{n}^Z)\in\mR^3$ for $n=1,2,\dots,N$. 
(This will be defined in Algorithm~\ref{alg:LLR-BP4}, referred to as \emph{MBP$_4$}.
A conventional LLR-BP$_4$ is a special case of Algorithm~\ref{alg:LLR-BP4} with $\alpha=1$.)
 There are two types of messages iteratively passed  on each edge $(m,n)$ connecting variable node~$n$ and check node~$m$.
 	Let $\cM(n)$ denote the set of neighboring check nodes of variable node $n$,
 	and $\cN(m)$ denote the set of neighboring variable nodes of check node $m$.
 	(In other words, $\cN(m)$ is the support of $S_{m}$.)
  We will simplify a notation $\cM(n)\setminus\{m\}$ as $\cM(n)\setminus m$.
A \textit{variable-to-check} message $\lambda_{S_{mn}}(\Gamma_{n\to m})$ from variable node $n$ to check node $m$ carries  the log-likelihood ratio that  $E_n$ commutes or anticommutes with $S_{mn}$, where $\Gamma_{n\to m}=(\Gamma_{n\to m}^X,\Gamma_{n\to m}^Y,\Gamma_{n\to m}^Z)$ is the LLR distribution of $E_n=I$ over $E_n=W$, for $W\in\{X,Y,Z\}$, according to the messages from the other nodes $m'\in\cM(n)\setminus m$ (cf.~\eq{eq:gamma_nm1}), and the function $\lambda_W:\mR^3\rightarrow \mR$ is defined as
	\begin{equation*} 
	\lambda_W(\gamma^X,\gamma^Y,\gamma^Z)\triangleq\ln \frac{1+ e^{-\gamma^{W}}}{   e^{-\gamma^{X}}+e^{-\gamma^{Y}}+e^{-\gamma^{Z}}-e^{-\gamma^{W}}   }.
	\end{equation*}
On the other hand, suppose that $S_{mn}\neq I$; then by \eq{eq:z_m}, we have a check bit relation
	$$ \langle E_n, S_{mn}\rangle= z_m +\sum_{n'\in\sN(m)\setminus n} \langle E_{n'}, S_{mn'}\rangle \mod 2. $$
Consequently a \textit{check-to-variable} message  $\Delta_{m\to n}$ from check node $m$ to variable node $n$  will tell us  the log-likelihood ratio of whether  $E_n$ commutes or anticommutes with $S_{mn}$. 
More precisely, as shown in \cite{KL21a}, we have
	\begin{align} \label{eq:Dmn_GF2}
	\Delta_{m\to n} =&(-1)^{z_m}\underset{n'\in\sN(m)\setminus n}{\boxplus} \lambda_{S_{mn'}}(\Gamma_{n'\to m}),
	\end{align}
where for a set of $k$ real scalars $a_1,a_2,\dots,a_k \in \mR$, the operation $\boxplus$ is defined by 
\begin{equation*} 
\overset{k}{\underset{i=1}{\boxplus}} \, a_i = 2\tanh^{-1} \left( \prod_{i=1}^k \tanh\frac{a_i}{2} \right). 
\end{equation*}
Then $\Gamma_{n}$ is updated according to  $\Delta_{m\to n}$ for all $m\in\cM(n)$ and the initial distribution of $E_n$ (cf.~\eq{eq:gamma_n}).

The BP algorithm iteratively updates the LLRs of the marginal distributions $\{\Gamma_{n}\}_{n=1}^N$ according to the passed messages on the Tanner graph. 
If the Tanner graph has no cycles, BP will compute 
the exact marginal distributions \cite{Pea88,Wib96,MMC98,AM00,KFL01}. 
If there are not many short cycles, the approximation is usually very good  \cite{Wib96,MMC98,AM00}.

Here we discuss a four-cycle example. 
Consider $S=\left[\begin{smallmatrix} X&Y\\ Z&Z \end{smallmatrix}\right]$ and $E=IZ$. 
Then $z=(1,0)$, and it is difficult for a parallel-scheduled BP to determine whether the solution is $ZI$ or $IZ$, 
so BP will oscillate between $II$ and $ZZ$.
Poulin and Chung suggested to do heuristic processes like random perturbation between BP iterations to have opportunity to find $ZI$ or $IZ$ \cite{PC08}.
Actually, a serial-scheduled BP with proper message magnitude can easily infer a degenerate error $\hat E=ZI$ without additional processes.
We refer the readers to \cite{KL20,KL20b,KL21a} for more details about  refined BP decoding algorithms for quantum codes,
where the techniques of \textit{message normalization} and \textit{message scheduling} are adopted. 

 \section{BP Decoding as Energy Minimization} \label{sec:bp_me}

A classical decoding can be considered as an energy minimization problem \cite{BB89}.
Given the error syndrome, the energy function is defined with respect to the parity checks so that each error vector matching the syndrome 
 can be considered as a local minimum of the energy function. 
An iterative decoding algorithm can be used to find a local minimum of the energy function with an initial point defined by the channel statistics \cite{LBB98}.
In particular, a decoding algorithm based on gradient optimization was proposed in \cite{LBB98}.

 We would like to characterize the energy function minimization problem for BP decoding of quantum codes.  
 The energy function is defined on $\mR^{3N}$, where each point represents the LLRs in \eq{eq:LLRs}.
 Instead of using only the energy defined by parity checks \cite{BB89,LBB98}, 
 we  introduce an additional term regarding the distance between a data point and the channel statistics.
 Then we explain why conventional BP fails to solve this energy function minimization  problem for quantum codes with high degeneracy.
 To solve the problem, we introduce our MBP$_4$.

 \subsection{Energy function of BP decoding} \label{sec:BP_J}
 Assume that qubit $n$ undergoes a Pauli error according to a distribution $(p_n^I, p_n^X, p_n^Y, p_n^Z)$, where  $p_n^I+ p_n^X+ p_n^Y+ p_n^Z=1$.
 Let $\Lambda_n=(\Lambda_n^X,\Lambda_n^Y,\Lambda_n^Z)\in\mR^3$ for $n=1,2,\dots, N$, where 
	\begin{equation} \label{eq:llr_p}
	\Lambda_n^X=\ln \frac{p_n^I}{p_n^X},\quad   \Lambda_n^Y=\ln \frac{p_n^I}{p_n^Y},\quad   \Lambda_n^Z=\ln \frac{p_n^I}{p_n^Z}.
	\end{equation}
 The   channel statistics vector $\Lambda=(\Lambda_1,\Lambda_2,\dots,\Lambda_N)\in \mR^{3N}$ is the channel  information  we have before decoding. 
 For depolarizing errors with rate $\epsilon$, we have $(p_n^I, p_n^X, p_n^Y, p_n^Z)=(1-\epsilon,\, \epsilon/3,\, \epsilon/3,\, \epsilon/3)$ and hence
 $\Lambda_n^X=\Lambda_n^Y=\Lambda_n^Z=\ln \frac{1-\epsilon}{\epsilon/3}$.

 Suppose that the syndrome vector $z=(z_1,\dots,z_M)\in \{0,1\}^M$ is determined from stabilizers $S_m= \otimes_{n=1}^N S_{mn}$  for $m=1,\dots,M$.
 Then we define the energy function of the decoding problem with respect to the stabilizers $\{S_m\}_{m=1}^M$ and the syndrome vector $z$ as  
\begin{equation}
 J(\Gamma)= J_{\text{D}}(\Gamma) + \eta {J}_{\text{S}}(\Gamma), \label{eq:energy}
\end{equation}
 where $\eta>0$ is a real scalar,
\begin{equation*} 
 J_{\text{D}}(\Gamma)= \frac{1}{2} \left\| \Gamma-\Lambda\right\|_2^2,
\end{equation*}
 and 
\begin{equation} \label{eq:energy2} 
 \small
 {J}_{\text{S}}(\Gamma)= -\sum_{m=1}^M  2\tanh^{-1} \left( (-1)^{z_m} \prod_{n\in \cN(m)} \tanh\left(  \frac{\lambda_{S_{mn}}( \Gamma_n)}{2} \right) \right). 
\end{equation} 
The value of $\eta$ does not matter in the following discussion and will be postponed to discuss in Appendix~\ref{sec:eta}.

The second term ${J}_{\text{S}}(\Gamma)$ measures the satisfaction of each check $S_m$, similar to the case in \cite[Sec.~IV]{LBB98}.
The additional term $J_{\text{D}}(\Gamma)$, which is convex in $\Gamma$, measures the distance between a point $\Gamma\in \mR^{3N}$ and the channel statistics $\Lambda$.
This is critical since the initial channel statistics $\Lambda$ affects the performance of BP \cite{HFI12,KL21a} and this should be reflected in the energy function.
An interpretation is that an logical error close to $N(\cS)\setminus\pm\cS$ has small $J_\text{S}$ but large $J_\text{D}$.

 \subsection{Gradient decent optimization} \label{sec:GD}

 Next we show that BP decoding is like a gradient decent minimization of the corresponding energy function \eq{eq:energy}.
 This provides a better understanding of how BP works.

One can easily verify that
$\nabla J = \left(	\frac{\partial J}{\partial\Gamma_1^X},\frac{\partial J}{\partial\Gamma_1^Y},\frac{\partial J}{\partial\Gamma_1^Z}, ~ 
					\dots, ~
					\frac{\partial J}{\partial\Gamma_N^X},\frac{\partial J}{\partial\Gamma_N^Y},\frac{\partial J}{\partial\Gamma_N^Z} \right)
$
has 
\begin{align}
 &\frac{\partial J}{\partial\Gamma_n^W} = \Gamma_n^W-\Lambda_n^W  \notag\\
 &\quad + \sum_{m\in\sM(n)\atop S_{mn}=W } \frac{ \eta g_{mn}( \Gamma) e^{-\Gamma_n^W}}{1+e^{-\Gamma_{n}^W}} \widetilde{\Delta}_{m\to n}  \label{eq:parJ}\\
 &\quad - \sum_{m\in\sM(n)\atop \langle W, S_{mn}\rangle=1 } \frac{ \eta g_{mn}( \Gamma) e^{-\Gamma_n^W}}{e^{-\Gamma_{n}^X}+e^{-\Gamma_{n}^Y}+e^{-\Gamma_{n}^Z}-e^{-\Gamma_{n}^{S_{mn}}}} \widetilde{\Delta}_{m\to n}, \notag
\end{align}
 where 
 	\begin{equation} \label{eq:gmn}
 	g_{mn}( \Gamma) = \frac{ 1-\tanh^2\frac{\lambda_{S_{mn}}(\Gamma_n)}{2} }{ 1-  \left(\prod_{l\in\sN(m)} \tanh\frac{\lambda_{S_{ml}}(\Gamma_l)}{2} \right)^2 } > 0
 	\end{equation}
 	and
 	\begin{align}
 	\widetilde{\Delta}_{m\to n} =&(-1)^{z_m}  \prod_{n'\in\sN(m)\setminus n} \tanh\frac{\lambda_{S_{mn'}}(\Gamma_{n'})}{2}. \label{eq:delta_tilde}
 	\end{align}
 	Notice that $\widetilde{\Delta}_{m\to n}$ in \eq{eq:delta_tilde} and $\Delta_{m\to n}$ in \eq{eq:Dmn_GF2} are similar but different in two aspects.
 	First, they have the same sign but $\Delta_{m\to n}$ is resized by $\tanh^{-1}$. 
	Second, $\Gamma_{n}$ used in \eq{eq:delta_tilde} and $\Gamma_{n\to m}$ used in \eq{eq:Dmn_GF2} 
	differ by a term $\langle W,S_{mn}\rangle \Delta_{m\to n}$ (cf.~\eq{eq:gamma_nm}).

 	The gradient decent method will update $\Gamma_{n}^W$ by
 	\begin{equation} \label{eq:GD}
 	\Gamma_{n}^W ~\leftarrow~ \Gamma_{n}^W- \zeta \frac{\partial J}{\partial\Gamma_n^W}, 
 	\end{equation}
 	where $\zeta>0$ is the decent step-size.  
	Herein, we consider a fixed decent step-size and simply assume $\zeta=1$.
 	Let $\omega_{mn}^{(0)}=  \frac{\eta g_{mn}( \Gamma) e^{-\Gamma_n^W}}{1+e^{-\Gamma_{n}^W}}$ and $\omega_{mn}^{(1)}=\frac{\eta g_{mn}( \Gamma) e^{-\Gamma_n^W}}{e^{-\Gamma_{n}^X}+e^{-\Gamma_{n}^Y}+e^{-\Gamma_{n}^Z}-e^{-\Gamma_{n}^{S_{mn}}}}$.
 	We have   $\Gamma_{n}^W$  updated as
 \begin{align}
 \Lambda_n^W- \sum_{m\in\sM(n)\atop S_{mn}=W } \omega_{mn}^{(0)} \widetilde{\Delta}_{m\to n}
 + \sum_{m\in\sM(n)\atop \langle W, S_{mn}\rangle=1 }\omega_{mn}^{(1)} \widetilde{\Delta}_{m\to n}. \label{eq:gamma_update2}
 \end{align}

 Since $\eta g_{mn}(\Gamma) > 0$, both $\omega_{mn}^{(0)}$ and $\omega_{mn}^{(1)}$ are positive.
 For simplicity, consider fixed $\omega_{mn}^{(0)}=\beta$ and $\omega_{mn}^{(1)}=\alpha^{-1}$.
 Consequently, the step update in \eq{eq:gamma_update2} depends on terms $\widetilde{\Delta}_{m\to n}$, especially those with large magnitude.

 BP has a similar computation, but with more elegant step update determined by terms $\Delta_{m\to n}$ (cf.~\eq{eq:gamma_n}),
 	because the magnitude of $\Delta_{m\to n}$ (which is resized by $\tanh^{-1}$) provides adequate (enough-large) strength for update.
 An evidence is that $\Gamma_n$ tends to have correct marginal distribution if updated by $\Delta_{m\to n}$, as shown in \cite{KL21a}.
 We conduct some pre-simulations for several codes and different $(\alpha,\beta)$, 
 and observe that updating $\Gamma_n$ (for decoding) is better using $\Delta_{m\to n}$ than $\widetilde{\Delta}_{m\to n}$.

 \subsection{Energy topology} \label{sec:J_S}

 Before we develop our \ourBP (in Sec.~\ref{sec:MBP}), let us first analyze the difficulties that BP suffers in decoding quantum codes with high degeneracy.

\begin{figure}
		\subfloat[\label{fig:BDD_cls}]        {\includegraphics[width=0.23\textwidth]{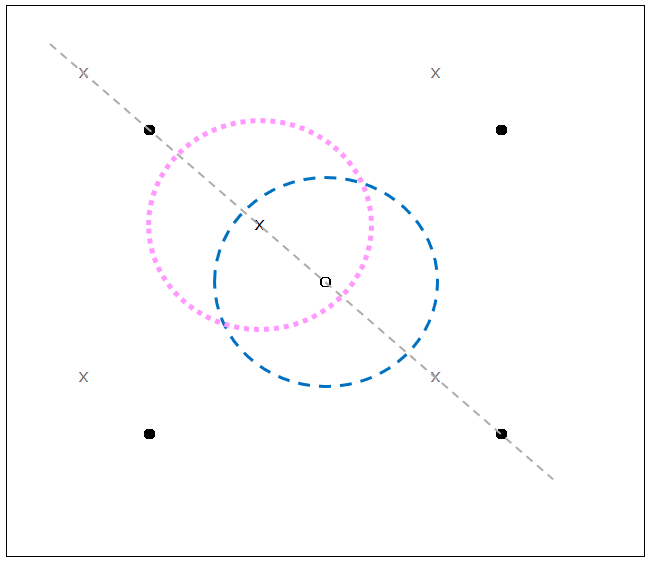}} \quad
		\subfloat[\label{fig:BDD_quan}]       {\includegraphics[width=0.23\textwidth]{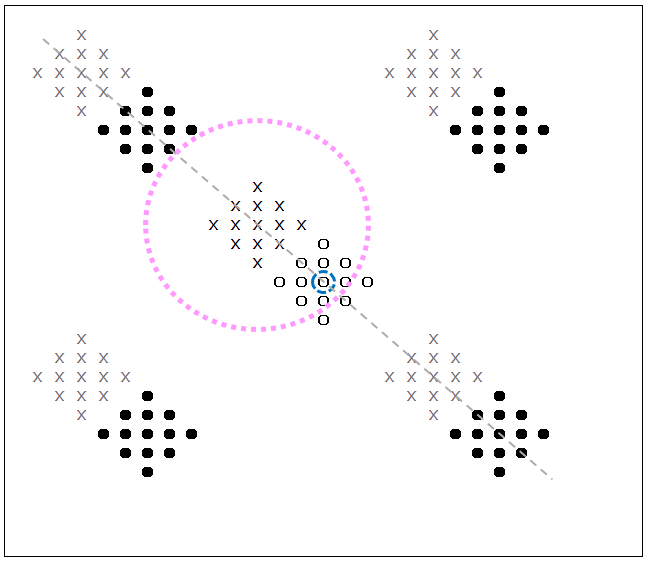}} \vfill
		\subfloat[\label{fig:BDD_cls_energy}] {\includegraphics[width=0.23\textwidth]{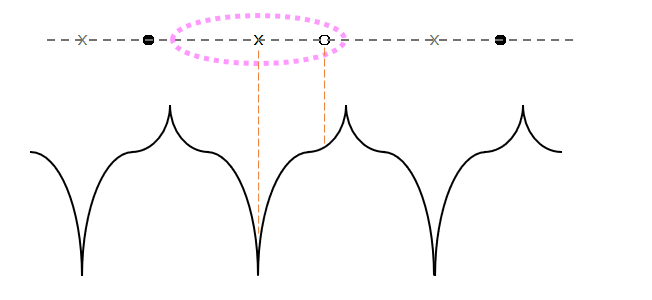}} \quad
		\subfloat[\label{fig:BDD_quan_energy}]{\includegraphics[width=0.23\textwidth]{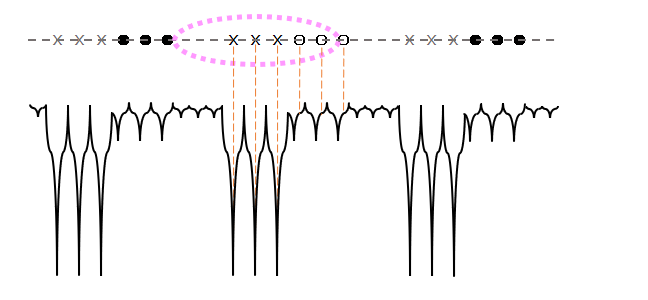}}
		\caption{
			Illustrations of the energy function $J_\text{S}$ of the decoding problem.
			(a) A classical code. 
			(b) A degenerate quantum code.
			(c)~and~(d) are the energy profiles along the dashed lines in (a) and (b), respectively.
		} \label{fig:BDD}	 
\end{figure}

  One may also consider the   energy function $J$ in \eq{eq:energy} as a sum of  the convex distance and the non-convex parity-check satisfaction terms. Since the non-convex term is the more difficult part in optimization,  for simplicity, we analyze the energy topology with only the parity-check satisfaction term ${J}_\text{S}(\Gamma)$.
We illustrate ${J}_\text{S}$ like a topography on $\Gamma$ for either a classical or a quantum decoding problem in Fig.~\ref{fig:BDD}.
Note that mathematically the function $J_\text{S}$ has several singular points corresponding to operators that match the given syndrome. 
However, BP will not get to these singular points since the values of the LLRs are  numerically protected in simulation.

First, consider a classical code with minimum distance $d$.
In Fig.~\ref{fig:BDD}\,(a), the small circle in the center denotes the all-zero vector, which will be called the \textit{origin} of the topography. 
The other black solid circles denote certain low-weight codewords.
The (blue) dashed circle is a classical {\it Hamming ball} with diameter $d$ centered at the origin.
The target error is denoted by a cross in the (blue) Hamming ball. 
The other crosses have the same error syndrome as the target error.
We also draw a (purple) dotted circle with diameter $d$ centered at the target error.
There will be energy barriers around this circle's boundary as shown in   Fig.~\ref{fig:BDD}\,(c), which is the  energy profile along the diagonal dashed line in Fig.~\ref{fig:BDD}\,(a). 
A syndrome-based BP starts from the origin and its goal is to find the target error, which corresponds to a global minimum of the energy function.
In our case, the starting point lies in a local convex hull of the target error. Since BP is like a gradient decent optimization as shown in the previous subsection, the target error can be found by BP.
Sometimes, the shape of a local minimum is very narrow and BP may converge better using a smaller step-size.
This is usually done by message normalization \cite{CF02a,CDE+05}.

Next, consider a nondegenerate quantum code with minimum distance $D=d$, where $d=\min\{\text{weight of } F: F\in N(\cS), F\ne \pm I^{\otimes N}\}$.
The decoding topography is like Fig.~\ref{fig:BDD_cls} and, indeed, 
BP usually works well with 
a small step-size (see examples in \cite{KL20,KL20b,KL21a}).

Finally, consider a degenerate quantum code with $D$ much larger than $d$.
Figure~\ref{fig:BDD}\,(b) illustrates the energy topography of this case. 
A set of small circles in the center denotes the low-weight stabilizers.
Each group of black solid circles denotes 
a set of equivalent logical operators in $N(\cS)\setminus\pm\cS$.
A classical Hamming ball becomes very small, as the (blue) dashed circle with diameter~$d$ centered at the origin ($I^{\otimes N}$).
This is because the quantum code is degenerate and there are low-weight stabilizers closer to $I^{\otimes N}$ than the other operators in  $N(\cS)\setminus\pm\cS$.
The target error and its low-weight degenerate errors are the crosses in the (purple) dotted circle with diameter $D$. The other groups of crosses are logical errors with the same error syndrome as the target error.  
Similarly, a syndrome-based BP starts from the origin.
The energy profile along the diagonal dashed line in Fig.~\ref{fig:BDD}\,(b) is shown in Fig.~\ref{fig:BDD}\,(d).
Because low-weight stabilizers are like low-weight classical codewords, there are many ripples in the shape of the topography, causing conventional BP to get trapped or wander around this region (or sometimes oscillate).

To help BP escape these local traps, one should use a large enough step-size, unlike the classical strategy, which favors a small step-size for convergence.
On the other hand, due to the degeneracy of the quantum code,  there are many equivalent solutions (such as the crosses in the (purple) dotted circle in Fig.~\ref{fig:BDD_quan}) and it suffices to find a degenerate error of the target.
Exploiting this degeneracy improves the decoding performance.
Using a large step-size helps BP to approach any of the degenerate errors. 
However, using only larger steps in BP may deteriorate the convergence behavior. 
In the next subsection we will show  that this issue can be mitigated by introducing a mechanism of \emph{inhibition}.

\subsection{BP with additional memory effects (MBP)}\label{sec:MBP}

Motivated from the analysis of the energy topology of a degenerate quantum code and the gradient decent optimization process in the previous subsections, we propose a quaternary BP in log domain with additional memory effects as follows.
To do an iterative update, consider an equation inspired from \eq{eq:gamma_update2} with $\widetilde{\Delta}_{m\to n}$ replaced by $\Delta_{m\to n}$ in \eq{eq:Dmn_GF2} and with fixed inhibition (inspired from BP) as
\begin{align}
\Gamma_{n\to m}^W =& \Lambda_n^W 
	+ \frac{1}{\alpha} \sum_{m'\in\sM(n) \atop \langle W, S_{m'n} \rangle=1} \Delta_{m'\to n} 
	- {\beta} \sum_{m'\in\sM(n) \atop   S_{m'n}= W } \Delta_{m'\to n} 
	\notag \\
	&- \langle W,S_{mn}\rangle \Delta_{m\to n}. \label{eq:gamma_nm2}
\end{align}

We notice that the term ${-\beta} \sum_{m'\in\sM(n) \atop   S_{m'n}= W } \Delta_{m'\to n}$ is from gradient decent optimization, but not in BP when approximating the marginal distribution \cite{KL21a}.
Furthermore, we find that in most cases, $\beta=0$ has better performance.\footnote{
	With the five-qubit code, better performance may be obtained by setting $\beta \ne 0$. This will be discussed in Sec.~\ref{sec:Rmk513}.
	}
Thus, instead of using \eq{eq:gamma_nm2}, we consider
\begin{align}
\Gamma_{n\to m}^W =& 
\Lambda_n^W + \frac{1}{\alpha} \sum_{m'\in\sM(n) \atop \langle W, S_{m'n} \rangle=1} \Delta_{m'\to n} - \langle W,S_{mn}\rangle \Delta_{m\to n}. \label{eq:gamma_nm1}
\end{align}

To compute the required quaternary distributions, we propose Algorithm~\ref{alg:LLR-BP4}, referred to as \ourBP.
Our update rule for $\Gamma_{n}$ in   \eq{eq:gamma_n} parallels to \eq{eq:gamma_update2}, suggested by the gradient of the energy topology, but with $\beta=0$ according to the previous discussion.

 	\begin{algorithm}
 		\begin{flushleft}
			\caption{: BP$_4$ with additional memory effects (\ourBP)} \label{alg:LLR-BP4}
 			\textbf{Input}: 
			$S \in\{I,X,Y,Z\}^{M\times N}$, $z \in\{0,1\}^M$,  $T_{\max}\in \mathbb Z_+$, \\ ~~~~ 
			real $\alpha >0$, and initial LLRs $\{(\Lambda_n^X, \Lambda_n^Y, \Lambda_n^Z) \in \mathbb R^3\}_{n=1}^N$.

 			{\bf Initialization.}  
 			For $n\in\{1,2,\dots,N\}$,  $m\in\sM(n)$,  and \\ ~~~~ 
			$W\in\{X,Y,Z\}$, let 
 			\begin{equation*}
 			\Gamma_{n\to m}^W = \Lambda_n^W.
 			\end{equation*}

 			{\bf Horizontal Step.} For $m\in\{1,2,\dots,M\}$ and $n\in\sN(m)$, \\ ~~~~ 
			compute
 			\begin{equation}
 			\Delta_{m\to n} = (-1)^{z_m}\underset{n'\in\sN(m)\setminus n}{\boxplus} \lambda_{S_{mn'}}(\Gamma_{n'\to m}). \label{eq:delta_mn}
 			\end{equation}
 			
 			{\bf Vertical Step.} For $n\in\{1,2,\dots,N\}$ and $W\in\{X,Y,Z\}$, \\ ~~~~ 
			compute
 			\begin{align}
 			\Gamma_{n}^W &= \Lambda_n^W + \frac{1}{\alpha} \sum_{m\in\sM(n) \atop \langle W, S_{mn} \rangle=1} \Delta_{m\to n}.
					\label{eq:gamma_n}
 			\end{align}

 			\begin{itemize}
 				\item ({\bf Hard Decision.}) 	Let $\hat E = \hat E_1\hat E_2\cdots\hat E_N$, where:\\
				$\hat E_n = I$,\, if $\Gamma_{n}^W > 0$ for all $W\in\{X,Y,Z\}$; or\\
 				$\hat E_n = \argmin\limits_{W\in\{X,Y,Z\}} \Gamma_{n}^W$, otherwise.
 			\end{itemize}

 			\begin{itemize}
 				\item If $\langle \hat E, S_m \rangle = z_m ~\forall~ m$, halt   and return ``CONVERGE'';
 				\item[-] Otherwise, if the maximum number of iterations $T_{\max}$ is reached, halt and return ``FAIL'';
 				\item[-] ({\bf Fixed Inhibition.}) Otherwise,
 				for $n\in\{1,2,\dots,N\}$,  $m\in\sM(n)$,  and $W\in\{X,Y,Z\}$, compute
 				\begin{align}
 				\Gamma_{n\to m}^W &= \Gamma_n^W -\ \langle W,S_{mn}\rangle \Delta_{m\to n}.\label{eq:gamma_nm}
 				\end{align}
 				\item[-] Repeat from the horizontal step.
 			\end{itemize}

 		\end{flushleft}
 	\end{algorithm}

\begin{remark} \label{rmk:inhibition}
 The presentation of Algorithm~\ref{alg:LLR-BP4} differs from that in \cite{KL21a}
 in a way that the term $- \langle W,S_{mn}\rangle\Delta_{m\to n}$, called \emph{inhibition}, is separated out.
  Unlike \cite{KL20,KL21a}, where the corresponding inhibition is scaled by $1/\alpha$, 
  we suggest to keep the inhibition strength  fixed, 
  since this part is the belief inherited in check node $m$ 
  and should not be altered when we update the outgoing belief in variable~$n$ 
  to make the decoding less affected by the short cycles.
\end{remark}

\begin{figure}
  	\centering \includegraphics[width=0.48\textwidth]{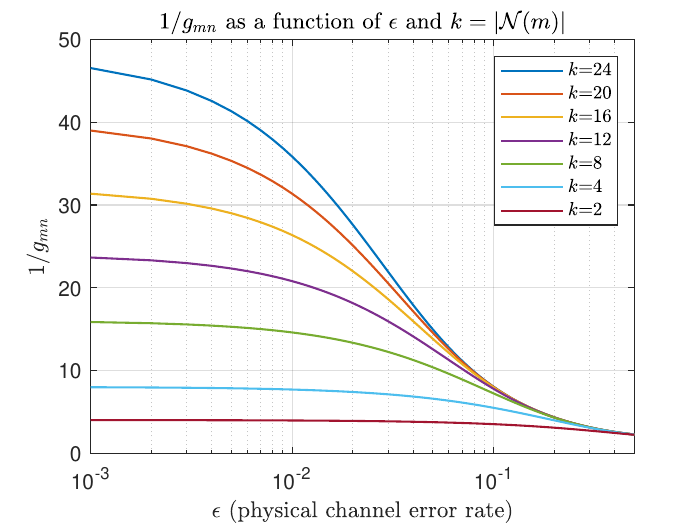}
  	\caption{
  		Plot of $1/g_{mn}$ (which is proportional to $\alpha$) as a function of $\epsilon$ for various row-weight ${k=|\sN(m)|}$.
  	} \label{fig:gmn_inv}				
   \end{figure}

 How to choose the factor $\alpha$ is intriguing.  
   The gradient optimization step suggests that  $\alpha \propto 1/g_{mn}(\Gamma)$ by \eq{eq:gamma_update2}. 
Since $\Gamma$ is initialized as the channel statistics vector $\Lambda$ at the first step and  $\alpha$ is fixed  in BP for simplicity, we 
   plot  $  1/g_{mn}(\Lambda )$ as a function of  the channel depolarizing rate $\epsilon$ in Fig.~\ref{fig:gmn_inv} for various stabilizer weight $k=|\cN(m)|$. 
The figure suggests two things.
  First,  $\alpha$  should be  larger as $\epsilon$  gets smaller. 
  Second, for a larger $k$, the maximum required $\alpha$ seems to saturate at a larger value. 
	The saturation suggests that the energy function becomes similar when $\epsilon$ get small enough, which means that there might be error-floor in the region of small $\epsilon$. However, this occurs at a smaller $\epsilon$ for a larger $k$.
   		These observations are consistent with our simulation results in Appendix~\ref{sec:bic2}, 
  in which we simulate as many values of $\alpha$ as possible for investigation.

\begin{remark} \label{rmk:cmplx}
A refined computation in Algorithm~\ref{alg:LLR-BP4} is:
	\begin{equation} \label{eq:gamma_nm3}
	\lambda_{S_{mn}} (\Gamma_{n\to m}) = \lambda_{S_{mn}} (\Gamma_n) -  \Delta_{m\to n}.
	\end{equation}
It is more efficient to update  $\lambda_{S_{mn}}(\Gamma_{n\to m})$ in this way since for each $n$, 
computing $\lambda_{S_{mn}} (\Gamma_n)$ needs at most three computations of $\lambda_{S_{mn}}(\cdot)$  for $S_{mn}\in\{X,Y,Z\}$; 
on the other hand, directly computing $\lambda_{S_{mn}} (\Gamma_{n\to m})$ needs $|\sM(n)|$ (usually $\ge 3$) computations of $\lambda_{S_{mn}}(\cdot)$.

The computation in the horizontal step can be simplified as in Remarks 1 and 4 of \cite{KL21a}.
Then the MBP$_4$ complexity is $O(Nj)$ per iteration. This is verified in Appendix~\ref{sec:runtime}.
\end{remark}

For reference, we provides  the  MBP$_4$ in linear domain  
in Appendix~\ref{sec:MBP_LD} to be compared with \mbox{\cite[Algorithm~3]{KL20}}.

\subsection{MBP decoding as an RNN} \label{sec:QuanBP}

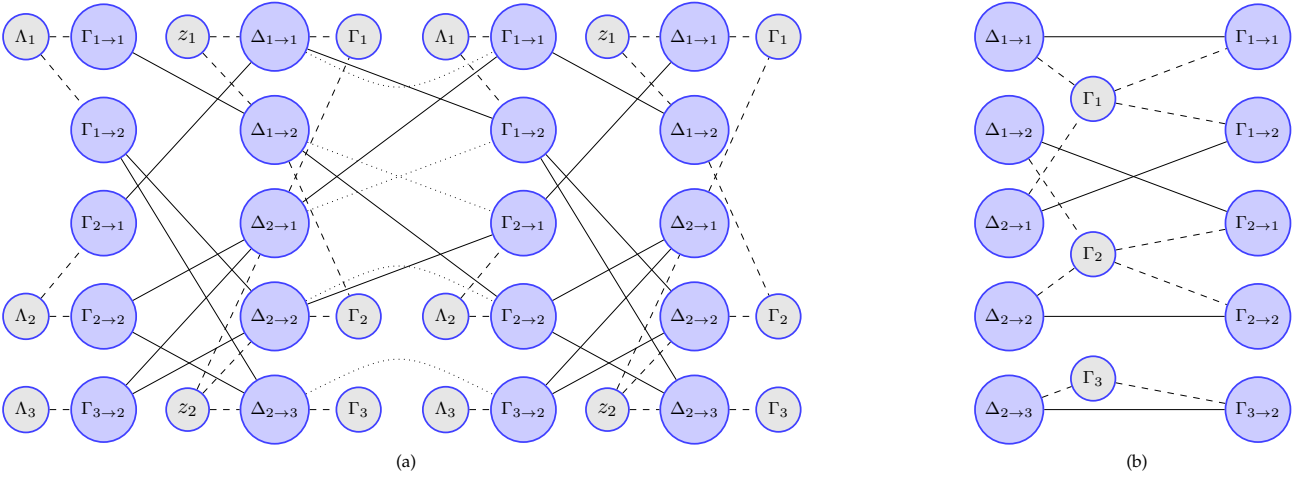
\begin{figure*}
	\resizebox{0.95\textwidth}{!}{ 
		\subfloat[\label{fig:S2x3_RNN_a}]{\begin{tikzpicture}[node distance=1.3cm,>=stealth',bend angle=45,auto]

\tikzstyle{chk}=[rectangle,thick,draw=black!75,fill=black!20,minimum size=4mm]
\tikzstyle{var}=[circle,thick,draw=blue!75,fill=gray!20,minimum size=4mm,font=\normalsize]
\tikzstyle{VAR}=[circle,thick,draw=blue!75,fill=blue!20,minimum size=5mm,font=\normalsize]
\tikzstyle{fac}=[anchor=west,font=\large]


\node[var] (p1) at (0.75,6.0) {\footnotesize $\Lambda_1$};
\node[var] (p2) at (0.75,1.5) {\footnotesize $\Lambda_2$};
\node[var] (p3) at (0.75,0.0) {\footnotesize $\Lambda_3$};

\node[VAR] (d11) at (2.0,6.0) {\footnotesize $\Gamma_{1\to 1}$};
\node[VAR] (d12) at (2.0,4.5) {\footnotesize $\Gamma_{1\to 2}$};
\node[VAR] (d21) at (2.0,3.0) {\footnotesize $\Gamma_{2\to 1}$};
\node[VAR] (d22) at (2.0,1.5) {\footnotesize $\Gamma_{2\to 2}$};
\node[VAR] (d32) at (2.0,0.0) {\footnotesize $\Gamma_{3\to 2}$};

\draw[dashed]	(p1) -- (d11);
\draw[dashed]	(p1) -- (d12);
\draw[dashed]	(p2) -- (d21);
\draw[dashed]	(p2) -- (d22);
\draw[dashed] 	(p3) -- (d32);

\node[VAR] (dt11) at (5.0-0.25,6.0) {\footnotesize $\Delta_{1\to 1}$};
\node[VAR] (dt12) at (5.0-0.25,4.5) {\footnotesize $\Delta_{1\to 2}$};
\node[VAR] (dt21) at (5.0-0.25,3.0) {\footnotesize $\Delta_{2\to 1}$};
\node[VAR] (dt22) at (5.0-0.25,1.5) {\footnotesize $\Delta_{2\to 2}$};
\node[VAR] (dt23) at (5.0-0.25,0.0) {\footnotesize $\Delta_{2\to 3}$};

\draw[thin]	(d21) -- (dt11);
\draw[thin]	(d11) -- (dt12);
\draw[thin]	(d22) -- (dt21);
\draw[thin]	(d32) -- (dt21);
\draw[thin]	(d12) -- (dt22);
\draw[thin]	(d32) -- (dt22);
\draw[thin]	(d12) -- (dt23);
\draw[thin]	(d22) -- (dt23);

\node[var] (z1) at (3.6-0.25,6) {$z_1$};
\node[var] (z2) at (3.6-0.25,0) {$z_2$};

\draw[dashed]	(z1) -- (dt11);
\draw[dashed]	(z1) -- (dt12);
\draw[dashed]	(z2) -- (dt21);
\draw[dashed]	(z2) -- (dt22);
\draw[dashed]	(z2) -- (dt23);

\node[var] (q1) at (6.35-0.25,6.0) {\footnotesize $\Gamma_1$};
\node[var] (q2) at (6.35-0.25,1.5) {\footnotesize $\Gamma_2$};
\node[var] (q3) at (6.35-0.25,0.0) {\footnotesize $\Gamma_3$};

\draw[dashed]	(dt11) -- (q1);
\draw[dashed]	(dt12) -- (q2);
\draw[dashed]	(dt21) -- (q1);
\draw[dashed]	(dt22) -- (q2);
\draw[dashed]	(dt23) -- (q3);

\node[VAR] (D11) at (10-1.25,6.0) {\footnotesize $\Gamma_{1\to 1}$};
\node[VAR] (D12) at (10-1.25,4.5) {\footnotesize $\Gamma_{1\to 2}$};
\node[VAR] (D21) at (10-1.25,3.0) {\footnotesize $\Gamma_{2\to 1}$};
\node[VAR] (D22) at (10-1.25,1.5) {\footnotesize $\Gamma_{2\to 2}$};
\node[VAR] (D32) at (10-1.25,0.0) {\footnotesize $\Gamma_{3\to 2}$};

\draw[thin]	(dt21) -- (D11);
\draw[thin]	(dt11) -- (D12);
\draw[thin]	(dt22) -- (D21);
\draw[thin]	(dt12) -- (D22);

\node[var] (P1) at (8.75-1.25,6.0) {\footnotesize $\Lambda_1$};
\node[var] (P2) at (8.75-1.25,1.5) {\footnotesize $\Lambda_2$};
\node[var] (P3) at (8.75-1.25,0.0) {\footnotesize $\Lambda_3$};
\draw[dashed]	(P1) -- (D11);
\draw[dashed]	(P1) -- (D12);
\draw[dashed]	(P2) -- (D21);
\draw[dashed]	(P2) -- (D22);
\draw[dashed] 	(P3) -- (D32);

\node[VAR] (DT11) at (13.0-1.5,6.0) {\footnotesize $\Delta_{1\to 1}$};
\node[VAR] (DT12) at (13.0-1.5,4.5) {\footnotesize $\Delta_{1\to 2}$};
\node[VAR] (DT21) at (13.0-1.5,3.0) {\footnotesize $\Delta_{2\to 1}$};
\node[VAR] (DT22) at (13.0-1.5,1.5) {\footnotesize $\Delta_{2\to 2}$};
\node[VAR] (DT23) at (13.0-1.5,0.0) {\footnotesize $\Delta_{2\to 3}$};

\draw[thin]	(D21) -- (DT11);
\draw[thin]	(D11) -- (DT12);
\draw[thin]	(D22) -- (DT21);
\draw[thin]	(D32) -- (DT21);
\draw[thin]	(D12) -- (DT22);
\draw[thin]	(D32) -- (DT22);
\draw[thin]	(D12) -- (DT23);
\draw[thin]	(D22) -- (DT23);

\node[var] (Z1) at (11.6-1.5,6) {$z_1$};
\node[var] (Z2) at (11.6-1.5,0) {$z_2$};

\draw[dashed]	(Z1) -- (DT11);
\draw[dashed]	(Z1) -- (DT12);
\draw[dashed]	(Z2) -- (DT21);
\draw[dashed]	(Z2) -- (DT22);
\draw[dashed]	(Z2) -- (DT23);

\node[var] (Q1) at (14.35-1.5,6.0) {\footnotesize $\Gamma_1$};
\node[var] (Q2) at (14.35-1.5,1.5) {\footnotesize $\Gamma_2$};
\node[var] (Q3) at (14.35-1.5,0.0) {\footnotesize $\Gamma_3$};

\draw[dashed]	(DT11) -- (Q1);
\draw[dashed]	(DT12) -- (Q2);
\draw[dashed]	(DT21) -- (Q1);
\draw[dashed]	(DT22) -- (Q2);
\draw[dashed]	(DT23) -- (Q3);

\draw[dotted]	(dt11)  .. controls (6.75,5) .. (D11);
\draw[dotted]	(dt12) -- (D21);
\draw[dotted]	(dt21) -- (D12);
\draw[dotted]	(dt22)  .. controls (6.75,2.5) .. (D22);
\draw[dotted]	(dt23)  .. controls (6.75,1) .. (D32);


\end{tikzpicture}} \qquad\qquad\qquad\qquad \subfloat[\label{fig:S2x3_RNN_b}]{\begin{tikzpicture}[node distance=1.3cm,>=stealth',bend angle=45,auto]

\tikzstyle{chk}=[rectangle,thick,draw=black!75,fill=black!20,minimum size=4mm]
\tikzstyle{var}=[circle,thick,draw=blue!75,fill=gray!20,minimum size=4mm,font=\normalsize]
\tikzstyle{VAR}=[circle,thick,draw=blue!75,fill=blue!20,minimum size=5mm,font=\normalsize]
\tikzstyle{fac}=[anchor=west,font=\large]





\node[VAR] (dt11) at (5.0-0.25,6.0) {\footnotesize $\Delta_{1\to 1}$};
\node[VAR] (dt12) at (5.0-0.25,4.5) {\footnotesize $\Delta_{1\to 2}$};
\node[VAR] (dt21) at (5.0-0.25,3.0) {\footnotesize $\Delta_{2\to 1}$};
\node[VAR] (dt22) at (5.0-0.25,1.5) {\footnotesize $\Delta_{2\to 2}$};
\node[VAR] (dt23) at (5.0-0.25,0.0) {\footnotesize $\Delta_{2\to 3}$};




\node[var] (q1) at (6.35-0.25,5.0) {\footnotesize $\Gamma_1$};
\node[var] (q2) at (6.35-0.25,2.5) {\footnotesize $\Gamma_2$};
\node[var] (q3) at (6.35-0.25,0.5) {\footnotesize $\Gamma_3$};

\draw[dashed]	(dt11) -- (q1);
\draw[dashed]	(dt12) -- (q2);
\draw[dashed]	(dt21) -- (q1);
\draw[dashed]	(dt22) -- (q2);
\draw[dashed]	(dt23) -- (q3);

\node[VAR] (D11) at (10-1.25,6.0) {\footnotesize $\Gamma_{1\to 1}$};
\node[VAR] (D12) at (10-1.25,4.5) {\footnotesize $\Gamma_{1\to 2}$};
\node[VAR] (D21) at (10-1.25,3.0) {\footnotesize $\Gamma_{2\to 1}$};
\node[VAR] (D22) at (10-1.25,1.5) {\footnotesize $\Gamma_{2\to 2}$};
\node[VAR] (D32) at (10-1.25,0.0) {\footnotesize $\Gamma_{3\to 2}$};

\draw[thin]	(dt11) -- (D11);
\draw[thin]	(dt12) -- (D21);
\draw[thin]	(dt21) -- (D12);
\draw[thin]	(dt22) -- (D22);
\draw[thin]	(dt23) -- (D32);

\draw[dashed]	(q1) -- (D11);
\draw[dashed]	(q1) -- (D12);
\draw[dashed]	(q2) -- (D21);
\draw[dashed]	(q2) -- (D22);
\draw[dashed]	(q3) -- (D32);

\end{tikzpicture}} 
	}
	\caption{
		(a) An RNN converted from Fig.~\ref{fig:S2x3}  with the first two iterations unrolled according to \eq{eq:gamma_nm1} and \eq{eq:delta_mn}. 
		Since there are five edges in Fig.~\ref{fig:S2x3}, the RNN has five neurons per hidden layer. 
		(b) Another (equivalent) way to update  $\Gamma_{n\to m}$ 
			between the two iterations in (a) according to \eq{eq:gamma_nm} (or \eq{eq:gamma_nm3}).
		In both subfigures, small circles are input/output neurons, and large circles are hidden neurons.
		A solid line is an edge connecting hidden neurons. 
		A dashed line is an edge connecting a hidden neuron and an input/output neuron.
		A dotted line in (a) is a special edge with weight $-(1-\frac{1}{\alpha})$ so that $\Delta_{m\to n}$ is rescaled and added to $\Gamma_{n\to m}^W$ 
		when $\langle W,S_{mn} \rangle=1$ as in \eq{eq:gamma_nm1}. 
		A such edge is implicitly embedded in (b) by \eq{eq:gamma_nm}. 
	} \label{fig:S2x3_RNN}
\end{figure*}

It was known that a BP decoding process can be modeled as an RNN \cite{Nac+18,LP19}.
Similarly we can derive an RNN from Algorithm~\ref{alg:LLR-BP4}. 
The RNN usually represents the BP with a parallel schedule.
In the Tanner graph induced by the ${M\times N}$ check matrix $S$,
two types of messages are iteratively updated: variable-to-check and check-to-variable messages.
Hence, there will be two hidden neuron layers computing messages $\Delta_{m\to n}$ or $\lambda_{S_{mn}}(\Gamma_{n\to m})$ per edge alternatively in each layer, 
and there are $N$ input neurons $\{\Lambda_n\}_{n=1}^N$ and $N$ output neurons $\{\Gamma_n\}_{n=1}^N$. 
($z_m$ can be considered embedded in $\Delta_{m\to n}$ or separated as an additional input neuron.)
An estimated error $\hat E$ will be inferred from $\{\Gamma_n\}_{n=1}^N$. 
The RNN will iterate until a valid $\hat E$ matching syndrome $z$ is found or a maximum number of iterations $T_{\max}$ is reached.

Figure~\ref{fig:S2x3_RNN_a} illustrates the RNN derived from the BP on the Tanner graph in Fig.~\ref{fig:S2x3}. 
A neuron denoted by $\Gamma_{n\to m}$ computes $\lambda_{S_{mn}} (\Gamma_{n\to m})$ although the symbol $\lambda_{S_{mn}}$ is not explicitly shown.
$\Gamma_n$ is updated also by $\Lambda_n$, although not shown in the figure.
Note that there are additional edges (dotted curves) from $\Delta_{m\to n}$ to $\Gamma_{n\to m}$, which
are not considered in the previous BP methods \cite{MMM04,PC08,PK19,RWBC20,KL20,KL20b,KL21a,GGKL21}, 
nor in the BP-based neural networks \cite{Nac+18,LP19}.

These (dotted) edges appear due to the fixed inhibition and provide additional memory effects, which contribute to the major improvement.
This agrees with the known result  that using proper inhibition between neurons enhances a network's perception capability in Hopfield nets \cite{Hop84,HT85,HT86,BM89,MWW91}.

Figure~\ref{fig:S2x3_RNN_b} shows  another (equivalent) way of updating $\Gamma_{n\to m}$ by \eq{eq:gamma_nm}.
According to the BP update rules,
every solid edge in Fig.~\ref{fig:S2x3_RNN_a} is an excitation (with positive weight);
in Fig.~\ref{fig:S2x3_RNN_b}, every dashed line is an excitation, but a solid line is an inhibition (with negative weight $-1$).

To sum up, we have shown that \ourBP can be extended to a neural network decoder, which may be improved by using appropriate weight for each edge.

\subsection{Adaptive MBP} \label{sec:AMBP}

\begin{algorithm}
\begin{flushleft}
{
	\caption{: Adaptive MBP$_4$ (AMBP$_4$): finding  $\alpha^*$ by incremental step-size scaled by $1/\alpha_i$, $i=1,2,\dots,\ell$.} \label{alg:AMBP}
		
	{\bf Input}:
	$S \in\{I,X,Y,Z\}^{M\times N}$, $z \in\{0,1\}^M$, $T_{\max}\in \mathbb Z_+$, 
	$\{\Lambda_n = (\Lambda_n^X, \Lambda_n^Y, \Lambda_n^Z) \in \mathbb R^3\}_{n=1}^N$, 
	a sequence of real values $\alpha_1 > \alpha_2 >\dots > \alpha_\ell > 0$, 
	and a function handler MBP$_4$.

	{\bf Initialization}: Let $i=1$. 

	{\bf MBP Step}: Run MBP$_4(S,\, z,\, T_{\max},\, \alpha_i,\, \{\Lambda_n\})$,
	\begin{itemize}
		\item[] which returns an indicator ``CONVERGE'' or ``FAIL'' with estimated $\hat E \in \{I,X,Y,Z\}^N$.
	\end{itemize}

	{\bf Adaptive Check}: 
	\begin{itemize}
		\item If the return indicator is ``CONVERGE'', return ``SUCCESS'' (with  valid $\hat E$ and $\alpha^* = \alpha_i$);
		\item Otherwise, let $i\leftarrow i+1$; if $i>\ell$, return ``FAIL''; 
		\item Otherwise, repeat from the MBP Step.
	\end{itemize}
}
\end{flushleft}
\end{algorithm}

In this subsection, we propose a variation of MBP with $\alpha$ chosen adaptively as shown in Algorithm~\ref{alg:AMBP}.
The value of $\alpha$ controls the search radius of MBP$_4$. Typically, a fixed $\alpha$ is chosen so that BP focuses on an error correction region, e.g., $1\times$BDD to $2\times$BDD. For codes with high degeneracy, we intend to correct errors of higher weight and consequently we need to consider variations in $\alpha$. 

Generating a solution by referring multiple decoding instances is an important technique in Monte Carlo sampling methods (cf.~parallel tempering in \cite{WL12}) and neural networks (cf.~\cite[Fig.~4]{Nac+18}). 
This is like an $\varepsilon$-net.
Thus we conduct multiple instances of MBP$_4$ 
each with a different value of~$\alpha$, and choose 
the solution from the instance with a largest $\alpha$ and valid solution.
This largest $\alpha$ is the most conservative $\alpha$ with a valid solution and will be denoted by $\alpha^*$. 
This adaptive scheme (Algorithm~\ref{alg:AMBP}) is referred to as AMBP$_4$.

Note that the procedure in Algorithm~\ref{alg:AMBP} tests each value of $\alpha_i$  in a sequential manner;
these $\alpha_i$'s can be tested in parallel  if the physical resources for implementation are available, followed by a final check to determine $\alpha^*$.

\section{Simulations Results} \label{sec:Sim}

In this section, we first study \ourBP on the well-known $[[5,1,3]]$ code \cite{LMPZ96}.
Then we simulate the decoding performance of (A)MBP$_4$ with various values of $\alpha$ on quantum bicycle codes, a GHP code,  and surface codes. 
(The results of toric codes are shown in Appendix~\ref{sec:toric}).

When evaluating the logical error rate or other error rates, an error bar between two crosses shows a 95\% confidence interval if fewer than 100 error events are collected.

In the following discussions, we will discard the global phase; or equivalently, when we refer $\cS$, it includes $\pm \cS$.

In simulations, a decoding is successful if it outputs an $\hat E\in E\sS$.
Thus it is important to determine whether $\hat E$ is a degenerate error of the target error $E$. 
For convenience, some literature may consider a decoding to be successful  only when $\hat E = E$ (such as in \cite{MMM04,Wan+12}); 
however, this is not accurate, especially for highly-degenerate codes.
A simple observation is that 
$\hat E\in E\sS$  if and only if $\hat{E}E \in\cS$. 
\begin{lemma}
Suppose that $\{S_m\}_{m=1}^{N-K}$ together with $\{\bar{X}_1,\bar{Z}_1,\bar{X}_2,\bar{Z}_2,\dots,\bar{X}_{K},\bar{Z}_{K}\}$ form a set of independent generators for $N(\cS)$. 
Then  $\hat E\in E\sS$ if and only if $\hat{E}E$ commutes with 
all elements in $\{S_m\}_{m=1}^{N-K} \cup \{\bar{X}_j,\bar{Z}_j\}_{j=1}^K$.
\end{lemma}
An efficient method to find $\{\bar{X}_j,\bar{Z}_j\}_{j=1}^K$ is by using the standard form discussed in \cite{CG96,GotPhD}.

Next we discuss how degeneracy is exploited. Let $n_{\rm tot}$ be the number of tested error samples for a data point in simulations.
Suppose that   $E^{(i)}$ and $\hat E^{(i)}$ are tested and estimated errors, respectively, for $i=1,2,\dots, n_{\rm tot}$.
Let   
   \begin{align}     	 
   n_{\rm 0}&= \mbox{\# of pairs } (E^{(i)},\hat E^{(i)}):  \hat{E}^{(i)}\ne E^{(i)} ,\\
   n_{\rm e}&= \mbox{\# of pairs } (E^{(i)},\hat E^{(i)}):  \hat E^{(i)}\notin E^{(i)}\sS ,\\
   n_{\rm u}&= \mbox{\# of pairs } (E^{(i)},\hat E^{(i)}):  \hat{E}^{(i)}E^{(i)} \in N(\cS)\setminus\cS .
   \end{align}
 Empirically, we have the \textit{classical block error rate} $P(\hat E\ne E) = n_{\rm 0}/n_{\rm tot}$, 
 the \textit{quantum logical error rate} $P(\hat E\notin E\sS) = n_{\rm e}/n_{\rm tot}$, and 
 the \textit{undetected error rate} $P(\hat E E\in N(\cS)\setminus \cS)=n_{\rm u}/n_{\rm tot}$.

Since  $(\hat E\notin E\sS) \subseteq (\hat E\ne E)$, by Bayes rule, we have
 \begin{equation} \label{eq:degen} 
 \begin{aligned}
 &\quad	P(\hat E\notin E\sS) = P(\hat E\notin E\sS, \hat E\ne E) \\
 & = P(\hat E\ne E)\times P(\hat E\notin E\sS \mid \hat E\ne E) = \frac{n_{\rm 0}}{n_{\rm tot}}\times\frac{n_{\rm e}}{n_{\rm 0}}.
 \end{aligned}
 \end{equation}

A classical strategy for improvement is trying to lower $n_{\rm 0}/n_{\rm tot}$,
which means that a target error needs to be accurately located for a given syndrome.
Such a strategy has a limit in performance due to short cycles or strong code degeneracy.
A better strategy should also try to lower $n_{\rm e}/n_{\rm 0}$, which will be called the error suppression ratio by exploiting degeneracy. For example, if the decoder converges to any of the degenerate errors, the decoding is a success (i.e., $n_{\rm e}$ does not increase although $n_{\rm 0}$ may increase by one).

 	Recall that  Algorithm~\ref{alg:LLR-BP4} with $\alpha=1$
 is equivalent to the conventional quaternary BP ({\it conventional BP$_4$}, or in short, {\it BP$_4$}). 
 We will demonstrate that MBP$_4$ outperforms BP$_4$ on various kinds of quantum codes.

For comparison, we will also consider  BP$_4$ with typical message normalization (denoted by $\alpha_c$) as in \cite{KL21a},
which is like the classical message normalization \cite{CF02a,CDE+05}.
Given some $\alpha_c>0$, consider that 
\eq{eq:gamma_nm} is replaced by
	\begin{align} 
	\Gamma_{n\to m}^W 
	&= \Lambda_n^W + \frac{1}{\alpha_c} \sum_{m'\in\sM(n) \atop \langle W, S_{m'n} \rangle=1} \Delta_{m'\to n} 
		- \frac{1}{\alpha_c}\langle W,S_{mn}\rangle \Delta_{m\to n} \notag \\
	&= \Lambda_n^W + \frac{1}{\alpha_c} \sum_{m'\in\sM(n)\setminus m \atop \langle W, S_{m'n} \rangle=1} \Delta_{m'\to n}, \label{eq:gamma_nmN}
	\end{align}
and this decoding will be referred to as \textit{normalized BP$_4$}. 
Notice how this is different from the rule \eq{eq:gamma_nm1} for MBP$_4$.

BP  can run with the parallel or serial schedule. (For the conversion of the schedules, see \cite{KL20}.)
Algorithm~\ref{alg:LLR-BP4} using each of these schedules will be referred to as {\it parallel MBP$_4$} or {\it serial MBP$_4$}. 
Similarly, we have  {\it parallel BP$_4$} or {\it serial BP$_4$} for conventional BP, 
and {\it parallel AMBP$_4$} or {\it serial AMBP$_4$} for Algorithm~\ref{alg:AMBP}. 
	We may also consider {\it parallel normalized BP$_4$} or {\it serial normalized BP$_4$}.
If a BP algorithm is referred without the prefix \emph{parallel} or \emph{serial}, the parallel schedule is assumed.

\subsection{The [[5,1,3]] code} \label{sec:Rmk513}

The $[[5,1,3]]$ code \cite{LMPZ96}  has a check matrix 
	$$S = \left[\begin{matrix}X&Z&Z&X&I\\ I&X&Z&Z&X\\ X&I&X&Z&Z\\ Z&X&I&X&Z\\\end{matrix}\right].$$
This code is worth investigation because there are many four-cycles in a small check matrix. 
We remark that using the serial schedule enables conventional BP$_4$ to decode the  $[[5,1,3]]$ code \cite{KL20}. 
Herein we use the parallel schedule to investigate the effect of $\alpha$, as well as $\beta$, in \eq{eq:gamma_nm2}. 

\begin{definition}
To understand the effect of $\beta$ in \eq{eq:gamma_nm2}, we consider an \emph{extended Algorithm~\ref{alg:LLR-BP4}} with 
an additional term $- {\beta} \sum_{m\in\sM(n) \atop   S_{mn}= W } \Delta_{m\to n}$ in the end of \eq{eq:gamma_n}.
\end{definition}

\begin{figure}
  	\centering \includegraphics[width=0.48\textwidth]{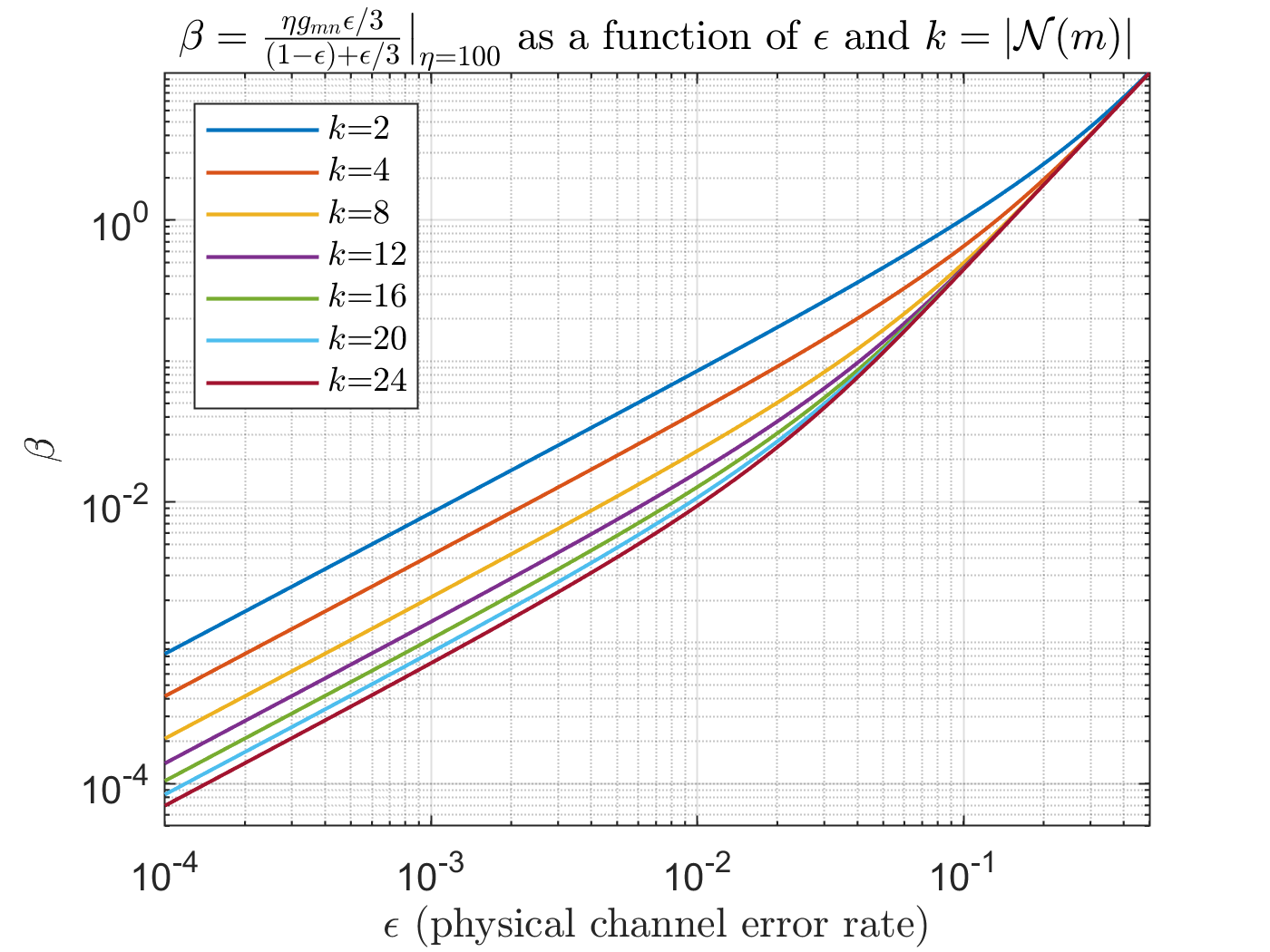}
  	\caption{
  		Plot of $\beta$ as a function of $\epsilon$ for various row-weight ${k=|\sN(m)|}$.
  	} \label{fig:beta}				
   \end{figure}

	According to \eq{eq:gamma_update2}, $\beta$ can be computed by $\omega^{(0)}_{mn}$ with ${\Gamma=\Lambda}$,  
	which suggests that $\beta = \frac{\eta g_{mn}(\Lambda) \epsilon/3}{(1-\epsilon)+\epsilon/3}$ with magnitude dominated by $\epsilon$.
	To see this, we set $\eta=100$ and plot $\beta$ as a function of $\epsilon$ in Fig.~\ref{fig:beta} for various row weight $k=|\cN(m)|$.

\begin{figure}
	\centering \includegraphics[width=0.5\textwidth]{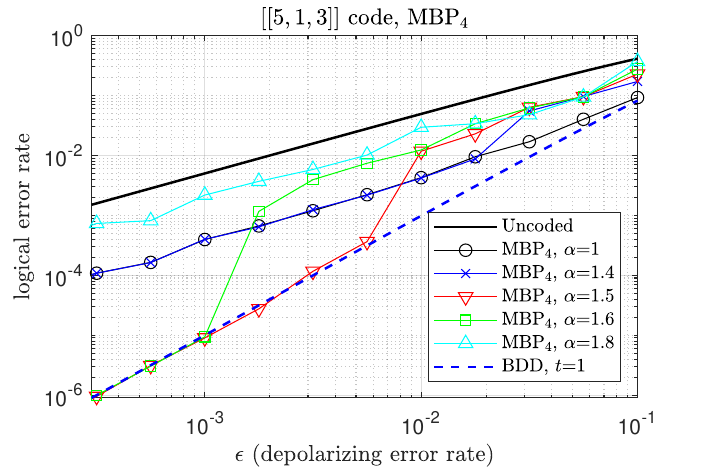}
	\caption{Decoding the $[[5,1,3]]$ code using Algorithm~\ref{alg:LLR-BP4} with various $\alpha$. 
	`Uncoded'' is uncoded block error rate of $N=5$.
	} \label{fig:dec513}	
\end{figure}

\begin{figure}
	\centering \includegraphics[width=0.5\textwidth]{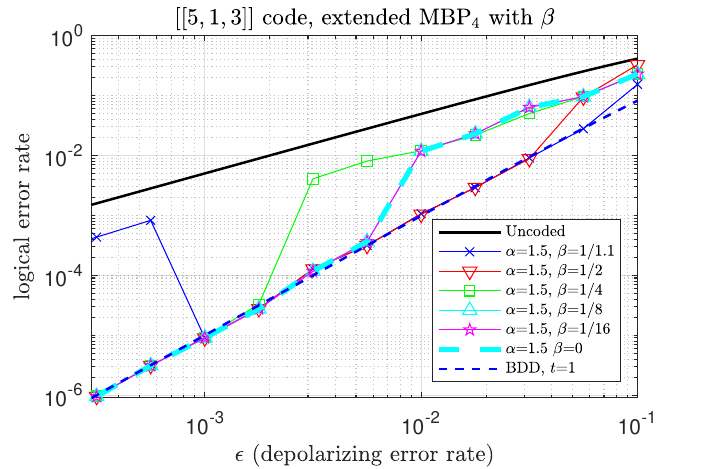}
	\caption{
		Decoding the $[[5,1,3]]$ code using extended Algorithm~\ref{alg:LLR-BP4} with ${\alpha=1.5}$ and various $\beta$. 
		`Uncoded'' is uncoded block error rate of $N=5$.
	} \label{fig:dec513b}
\end{figure}

We simulate Algorithm~\ref{alg:LLR-BP4} (i.e., $\beta=0$) with various $\alpha$ to decode the $[[5,1,3]]$ code. 
The performance curves 
are plotted in Fig.~\ref{fig:dec513}, including the BDD performance.
As can be seen, $\alpha \approx 1.5$ has the best performance, comparable with BDD at small $\epsilon$.
Next we simulate extended Algorithm~\ref{alg:LLR-BP4} with $\alpha=1.5$ and various $\beta$, and the results are plotted in Fig.~\ref{fig:dec513b}.
Observe that  large $\epsilon$ needs large $\beta$  and  small $\epsilon$ needs small $\beta$,  matching the observation in Fig.~\ref{fig:beta}.


The $[[5,1,3]]$ code can correct any single-qubit error, but conventional BP$_4$ (the case with $\alpha=1$) 
fails to decode the error $IIIYI$, no matter what value $\beta$ is. 

\begin{figure}
	\subfloat[\label{fig:Y4_ai=1}]{\includegraphics[height=0.136\textwidth]{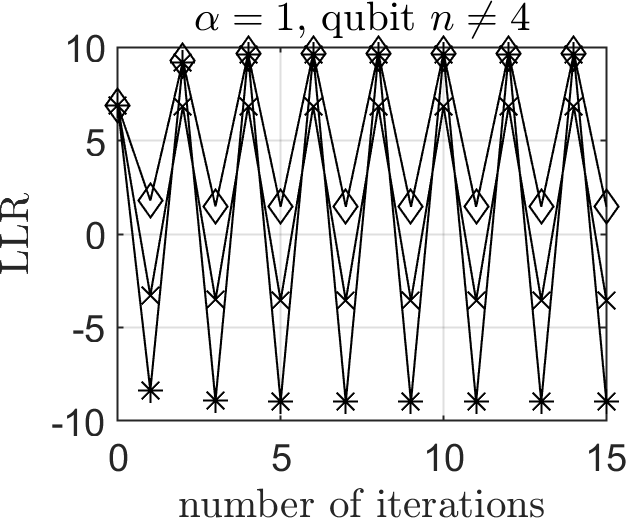}} ~~ \subfloat[\label{fig:Y4_ai=1.5}]{\includegraphics[height=0.136\textwidth]{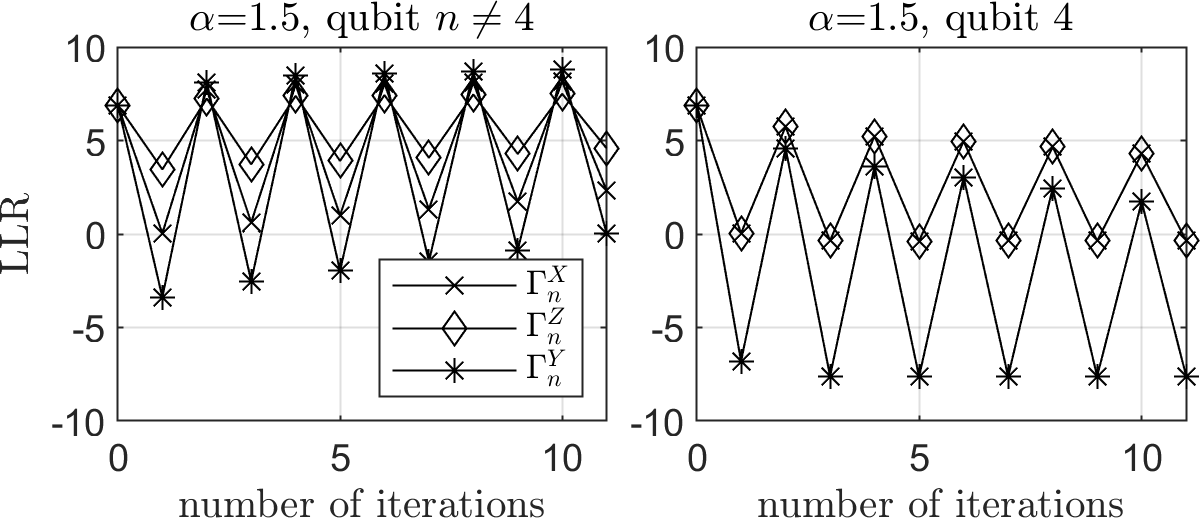}} 
	\caption{
		The change of the state $\Gamma_n=(\Gamma_n^X,\Gamma_n^Y,\Gamma_n^Z)$ in decoding the $[[5,1,3]]$ code with error $IIIYI$ at $\epsilon = 0.003$. 
		(a)~MBP$_4$ with $\alpha=1$ (conventional~BP$_4$) for qubit $n\ne 4$. The case of  $n=4$ is similar  but with a different amplitude.  
		(b)~MBP$_4$ with $\alpha = 1.5$.
	} \label{fig:Y4_ai}		\vspace*{\floatsep}
	\subfloat[\label{fig:chk_Y4}]{\includegraphics[height=0.138\textwidth]{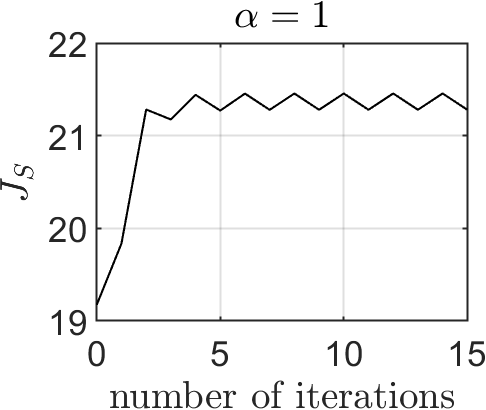}}
	\subfloat[\label{fig:chk_Y4_a15}]{\includegraphics[height=0.138\textwidth]{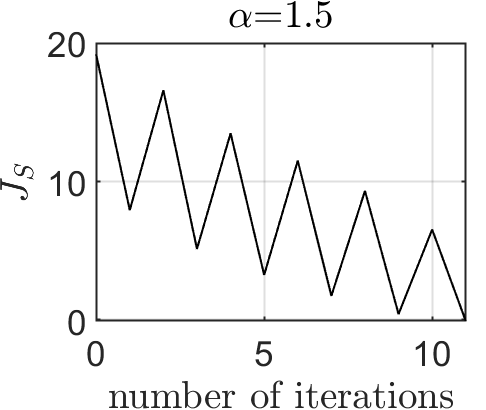}}
	\subfloat[\label{fig:chk_Y4_ac15}]{\includegraphics[height=0.138\textwidth]{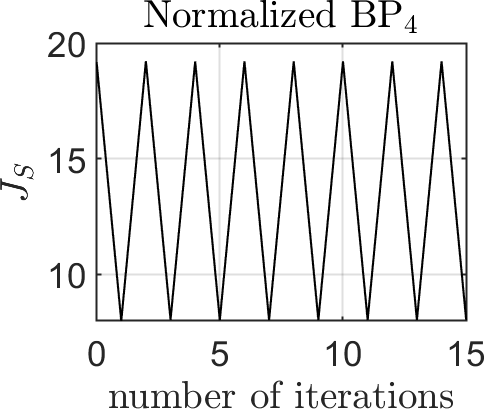}} 
	\caption{
		The evolution of $J_{\text{S}}$ in \eq{eq:energy2} for decoding the $[[5,1,3]]$ code with error $IIIYI$ at $\epsilon = 0.003$.
		(a)~MBP$_4$ with $\alpha=1$. 
		(b)~MBP$_4$ with $\alpha = 1.5$.
		(c)~Normalized BP$_4$ with $\alpha_c = 1.5$.
	} \label{fig:dm_ai}		
\end{figure}

In the following, we simply consider $\beta=0$.
We plot  $\Gamma_n$ at each iteration in decoding $IIIYI$ at $\epsilon = 0.003$ in Fig.~\ref{fig:Y4_ai}, 
where $\Gamma_n=\Lambda_n$ at iteration 0. 
With $\alpha=1$, the state of $\Gamma_n$ oscillates between $IIIII$ and $YYYYY$ continuously.
If $\alpha= 1.5$ is used instead,  \ourBP has a resistance to wrong beliefs, and the decoding converges correctly.

We plot the change of $J_{\text{S}}$ in Fig.~\ref{fig:dm_ai}. 
Figures~\ref{fig:dm_ai}\,(a) and \ref{fig:dm_ai}\,(b) correspond to Figs.~\ref{fig:Y4_ai=1} and \ref{fig:Y4_ai=1.5}, respectively.
With $\alpha=1$, the value of $J_{\text{S}}$ oscillates around $21.3$ with a small swing. 
Using $\alpha=1.5$, \ourBP enlarges the swing and lowers  $J_{\text{S}}$ to finally go to a small value $<0$.
This is because wrong beliefs are resisted and there are additional memory effects between iterations.
We also consider normalized BP$_4$ with \mbox{$\alpha_c=1.5$} as in \eq{eq:gamma_nmN}.
In this case, the swing is large but there are no memory effects, so it continuously oscillates between 8 and 19 and cannot have $J_{\text{S}}<0$, as shown in Fig.~\ref{fig:dm_ai}\,(c).

\begin{figure}
	\centering \includegraphics[width=0.5\textwidth]{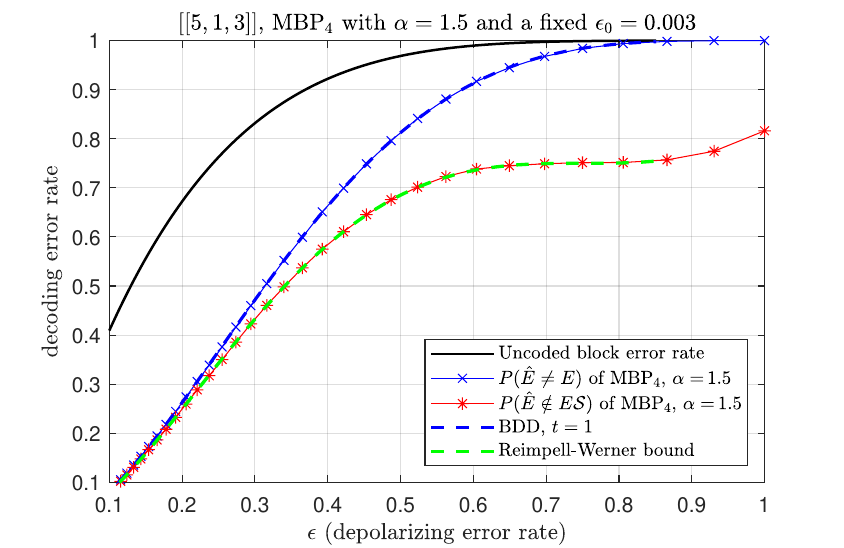}
	\caption{
		Decoding the $[[5,1,3]]$ code using parallel MBP$_4$ with $\alpha=1.5$ and a fixed $\epsilon_0=0.003$ to initialize $\Lambda_n$.
		$P(\hat E \ne E)$ is the classical block error rate, which is comparable with BDD.
		$P(\hat E \notin E\sS)$ is the quantum logical error rate, which matches the Reimpell-Werner   bound.
	} \label{fig:513_deg}
\end{figure}

Next we consider BP decoding with a fixed $\epsilon_0 = 0.003$ to initialize $\Lambda_n$, regardless of the actual depolarizing rate.
We find that the performance curve of \ourBP with $\alpha=1.5$ matches the BDD curve for small $\epsilon$.
Even more, it matches the Reimpell-Werner bound for the $[[5,1,3]]$ code \cite{RW05}, which outperforms the BDD at a large $\epsilon$ due to degeneracy.
The curves of BDD,  Reimpell-Werner bound, and \ourBP with $\alpha=1.5$ are compared in Fig.~\ref{fig:513_deg}.

\begin{remark} \label{rmk:fix_eps_0}
	Initializing $\{\Lambda_n\}$ with respect to a fixed $\epsilon_0$, instead of the actual depolarizing rate, is a strategy to maintain the decoding stability without  curve fluctuation \cite{HFI12}.\footnote{
		Sometimes we encounter the curve fluctuation, as in Figs.~\ref{fig:dec513} or~\ref{fig:dec513b}. 
		Using a fixed $\epsilon_0$ for initialization can improve this. 
		This can be proved by the theory in \cite{HFI12}.
		Put it more simply: if BP has $r\times$BDD performance at some physical error rate $\epsilon_0$, 
		then most errors within the corresponding correction radius would be correctable by BP. 
		Using an initialization with fixed $\epsilon_0$ enables BP to
		have the performance curve interpolated or extrapolated without fluctuation for different $\epsilon$. 
		}
	This also reflects the importance of choosing the network initial state \cite{SMDH13}.
	This technique of fixed initialization works for any quantum codes and is useful when the channel parameter is hard to estimate. 
\end{remark}


\subsection{Bicycle code} \label{sec:bic}

Bicycle codes are a kind of sparse-graph quantum codes with flexible length, code rate, and row-weight \cite{MMM04}. 
Let $k$ be the row-weight of a bicycle code. 
A smaller $k$ means the code has many low-weight stabilizers and the code is more degenerate. 
However, since the minimum distance $D\le k$ due to the construction, if $k$ is too-small, the code may have a high error-floor.

	\begin{figure}
	\subfloat[\label{fig:3786_Hk}]{\includegraphics[width=0.5\textwidth]{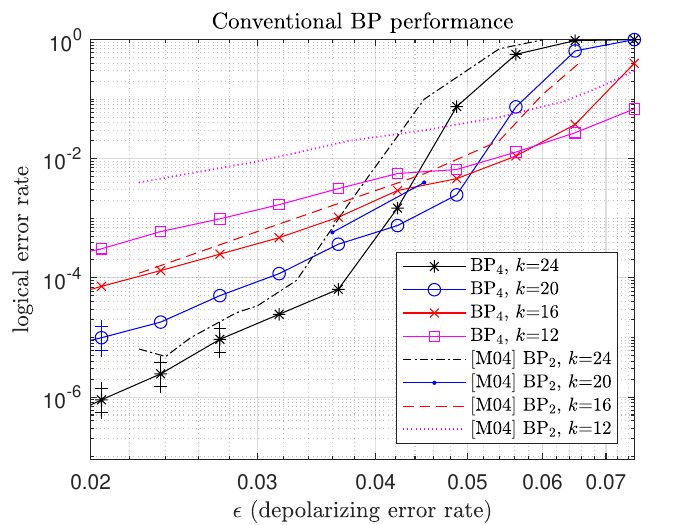}}\\
	\subfloat[\label{fig:3786_ai}]{\includegraphics[width=0.5\textwidth]{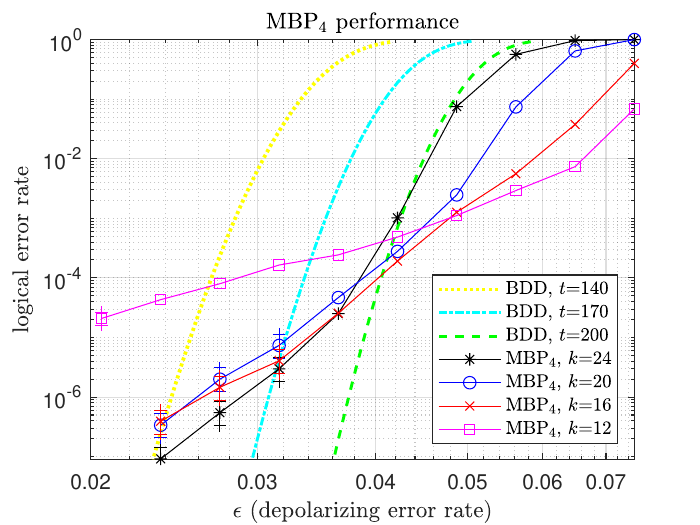}}\\
	\subfloat[\label{fig:3786_it}]{\includegraphics[width=0.5\textwidth]{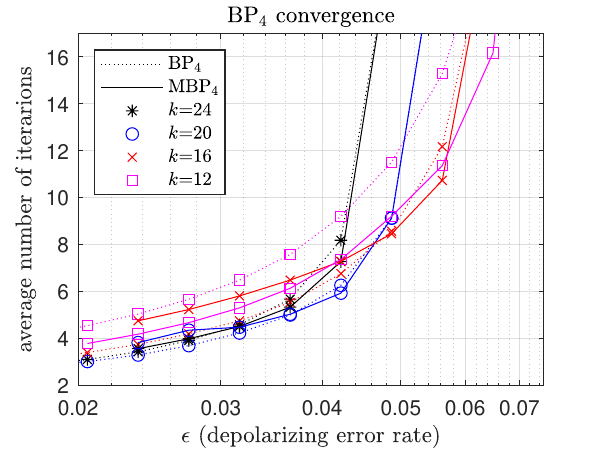}} 
	\caption{
		Performances of parallel BP$_4$ and MBP$_4$ on $[[3786,946]]$ bicycle codes with different row-weights $k$, based on $T_{\max}=90$. 
		(a)~MBP$_4$ with $\alpha=1$ (conventional BP$_4$). 
		(b)~MBP$_4$ with appropriate $\alpha>1$. 
		(c)~Average numbers of iterations in (a) (dotted lines) and (b) (solid lines).
		{The [M04] curves in (a) are from \cite[Fig.~6]{MMM04} (through a conversion from bit-flip error rate to depolarizing error rate \cite[Eq.~(40)]{MMM04}).}
	} \label{fig:3786}
	\end{figure}

\mbox{MacKay \it et al.} showed that, for a bicycle code with parameters $[[N,K]]=[[3786,946]]$, 
a row-weight $k\ge 24$ is required to have good binary BP (BP$_2$) performance  at a target block error rate $10^{-4}$ (see \cite[Fig.~6]{MMM04}).
We construct bicycle codes with the same parameters.
Figure~\ref{fig:3786}\,(a) shows the conventional BP$_4$ performance on $[[3786,946]]$ bicycle codes with $k=24,20,16,12$. 
It agrees with \cite[Fig.~6]{MMM04} that $k=24$ is required  to achieve the logical error rate of $10^{-4}$ before hitting the error-floor.
	Also shown in Fig.~\ref{fig:3786}\,(a) are the BP$_2$ performance curves in \cite{MMM04}.
	It can be seen that BP$_4$ performs better than BP$_2$, because the correlations between $X$ errors and $Z$ errors are considered in BP$_4$. 
	(The same phenomenon is discussed in~\cite[Fig.~11]{MMM04}.)

Now we use \ourBP with $\alpha>1$, and the performance is significantly improved, as shown in Fig.~\ref{fig:3786}\,(b).
(Note that different $\epsilon$ need different $\alpha$ as shown in Appendix~\ref{sec:bic2}.)
It shows that a bicycle code with row-weight $k=16$ is able to achieve the logical error rate of $10^{-6}$, which is less affected by the error-floor.
Thus we have greatly improved the BP performance on bicycle codes.

The minimum distance $D$ of a bicycle code is unknown, so it is hard to compare with $r\times$BDD as in Definition~\ref{def:BDD}.
Instead, we directly specify the correction radius $t$ and plot some BDD curves in Fig.~\ref{fig:3786}\,(b).
For $k=16$, depending on the logical error rate, the performance of MBP$_4$ is close to BDD with $t$ between 140 and 200, despite the fact that there is a far more smaller $D\le k=16$.
If $t=170$ is considered, since $D/2 \le 8$, it is better than 20$\times$BDD.

The average numbers of iterations are shown in Fig.~\ref{fig:3786}\,(c). 
	Fewer iterations are required if the decoder has better convergence behavior.
	After applying $\alpha$, the case of $k=12$ has obvious improvement, since there are more lower-weight stabilizers, i.e., there are more low-weight degenerate errors for MBP$_4$ to converge.
On the other hand, the cases of $k>12$ may require slightly more iterations to have MBP$_4$ improved from BP$_4$.

Although it is not shown, normalized BP$_4$ (cf.~\eq{eq:gamma_nmN}) also improves from BP$_4$, 
but MBP$_4$ outperforms normalized BP$_4$, especially for small $k$.
Normalized BP$_4$ can compete the neural BP \cite{LP19} on $[[256,32]]$ bicycle codes, as shown in \cite{KL21a}.
It would be difficult to train a neural BP on large code sizes such as $[[3786,946]]$ here.

\begin{table}
	\caption{
		Numbers of various events in the simulations of BP$_4$ and MBP$_4$ on the bicycles codes with row-weights 16 and 12.
	} 
	\label{tbl:bic_degen} \centering 
	$\begin{array}{|r|r|r|r|}
	\hline
	\text{BP$_4$ at $\epsilon$}:	& 0.027		& 0.037		& 0.049		\\
	\hline
	k=16:~n_{\rm tot}	& 398936	& 95977	& 21590		\\
	n_{\rm 0}~		& 100		& 100		& 103		\\
	n_{\rm e}~		& 100		& 100		& 100		\\
	n_{\rm u}~		& 0			& 0			& 0			\\
	\hline
	k=12:~n_{\rm tot}	& 101997	& 31765	& 15155		\\
	n_{\rm 0}~		& 134		& 154		& 206		\\
	n_{\rm e}~		& 100		& 100		& 100		\\
	n_{\rm u}~		& 1			& 0			& 3			\\
	\hline
	\hline
	\text{MBP$_4$ at $\epsilon$}:	& 0.027		& 0.037		& 0.049		\\
	\hline
	k=16:~n_{\rm tot}	& 12635150	& 3920434	& 79172		\\
	n_{\rm 0}~		& 34		& 126		& 102		\\
	n_{\rm e}~		& 20		& 100		& 100		\\
	n_{\rm u}~		& 0			& 0			& 0			\\
	\hline
	k=12:~n_{\rm tot}	& 1251769	& 410670	& 89821		\\
	n_{\rm 0}~		& 518		& 466		& 270		\\
	n_{\rm e}~		& 100		& 100		& 100		\\
	n_{\rm u}~		& 1			& 3			& 2			\\
	\hline
	\end{array}$
\end{table}

Next we study whether MBP$_4$ has good error suppression ratio $n_{\rm e}/n_{\rm 0}$ \eq{eq:degen}.
This ratio reduces if the decoder finds degenerate errors often when the actual error is not located. 
(M)BP$_4$ can have $n_{\rm e}/n_{\rm 0}<1$ for $k=16$ and $12$ in the region of $\epsilon\leq 0.049$.
Detailed event counts $n_{\rm 0}$, $n_{\rm e}$, $n_{\rm u}$, and $n_{\rm tot}$ in this region
are provided in Table~\ref{tbl:bic_degen}.
Observe that a decoder (especially MBP$_4$) exploits the degeneracy more for a code with smaller $k$ (stronger degeneracy). 
When $\epsilon$ gets smaller, the ratio $n_{\rm e}/n_{\rm 0}$ reduces significantly for MBP$_4$ on both codes, 
especially for $k=12$, although the minimum distance of this case is too small to have a low error-floor.
We remark that conventional BP$_4$ has $n_{\rm e}/n_{\rm 0}\approx 1$ for $k\ge 16$.
Also listed in Table~\ref{tbl:bic_degen} are  the numbers of undetected errors,
which are nonzero for ${ k=12 }$. However, the ratio $n_{\rm u}/n_{\rm tot}$ tends to be small.

\begin{figure}
	\centering \includegraphics[width=0.5\textwidth]{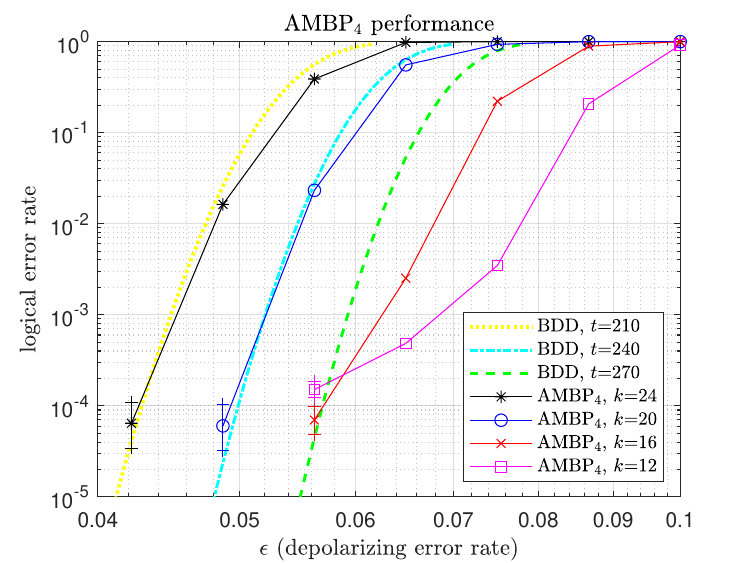}
	\caption{
	Performances of serial AMBP$_4$ on the $[[3786,946]]$ bicycle codes with different row-weights ($k$). 
	} \label{fig:AMBP_bic}		
\end{figure}

	To further improve the decoding performance, we apply AMBP$_4$ with  $\alpha^*\in\{2.4, 2.39, \dots, 0.5\}$. 
	Herein we consider the serial schedule because it accelerates the message update and enlarges the error-correction radius in finite iterations. 
		The performance curves in Fig.~\ref{fig:3786}\,(b) are significantly improved as shown in Fig.~\ref{fig:AMBP_bic}.

For quantum communication, we may consider a target logical error rate $10^{-4}$ \cite{MMM04}, and quantum retransmission is possible if necessary \cite{YLZ21}.
Consider that $\epsilon \approx t/N$ for large~$N$.
The quantum Gilbert--Varshamov rate \cite{EM96,GotPhD} states that there exist codes of rate $1/4$ to achieve arbitrarily low logical error rate at $\epsilon=0.063$ for asymptotically large $N$.
As shown in  Fig.~\ref{fig:AMBP_bic}, 
the $[[3786,946]]$ bicycle code with ${k=16}$ achieves logical error rate $10^{-4}$ at  $\epsilon=0.057$, which is close to the quantum Gilbert--Varshamov rate.

\subsection{Generalized hypergraph-product code} \label{sec:GHP}

\begin{figure}
	\centering 
	\includegraphics[width=0.5\textwidth]{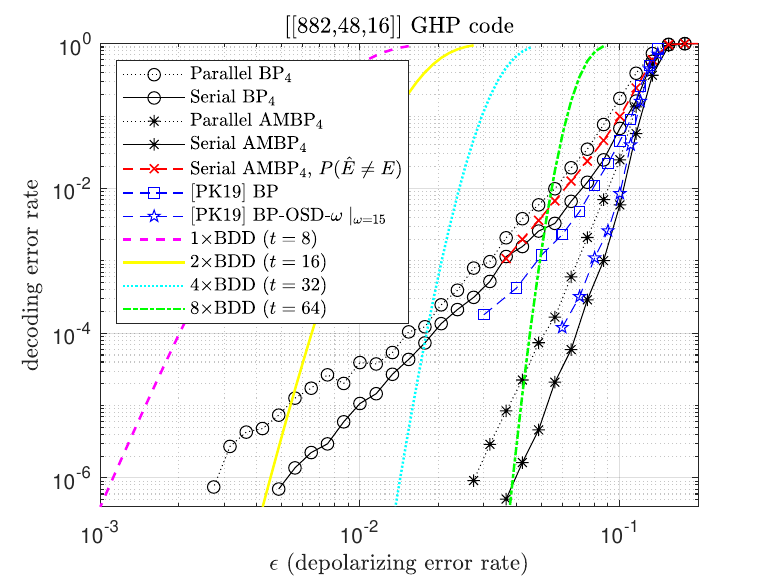}
	\caption{
	Performances of various BP on the $[[822,48,16]]$ GHP code, based on $T_{\max}=32$.
	The curve $P(\hat E\ne E)$ is the classical block error rate, while the other non-BDD curves are the quantum logical error rate with degenerate errors considered.
	The curves [PK19] are from \cite{PK19}.
	} \label{fig:GHP}		
\end{figure}

Herein we consider GHP codes \cite{KP13}, 
in particular, the $[[N,K,D]]= [[882,48,16]]$ GHP code constructed in \cite{PK19}, which has row-weight $8 < D$ and is thus highly-degenerate.
The performance of this code is shown in Fig.~\ref{fig:GHP}.
Using conventional BP$_4$ does not have good enough performance.
Using MBP$_4$, we find that most errors can be decoded with $\alpha\approx 1.2$ to $1.5$. 
However, since this code exhibits high degeneracy, a smaller $\alpha$ (larger step-size) may be needed.
Thus we apply  AMBP$_4$ with   $\alpha^* \in \{1.5, 1.49, \dots, 0.5\}$.

We use $r\times$BDD for reference. 
Observe that AMBP$_4$ has slope roughly aligned with $1\times$BDD or $2\times$BDD, but its performance is close to $8\times$BDD at logical error rate $10^{-6}$, since more low-weight errors are corrected.
We also draw the curve of  classical block error rate    $P(\hat E\ne E)=n_{\rm 0}/n_{\rm tot}$, which shows that the improvement of AMBP$_4$ from BP$_4$ mostly comes from exploiting degeneracy.

For reference, we also plot the performance curves given in \cite{PK19}, 
where the improved one is based on a layered-scheduled BP with post-processing by OSD-$w$, i.e, 
OSD is combined with a post-selection with a parameter $\omega$ to sort out $2^\omega$ errors in $\omega$ unreliable coordinates (so the overall complexity is high).
For most codes in \cite{FL95}, using $\omega=0$ is sufficient, but this GHP code needs large $\omega=15$ to achieve better performance.
As can be seen, AMBP$_4$ can outperform BP-OSD-15 on this code.
The complexity of AMBP$_4$ is low enough so we simulate to lower logical error rate.

\subsection{Surface code} \label{sec:surf}

In this subsection we consider the surface codes with a 45$^\circ$ rotation \cite{HFDM12}. A parallel study of rotated toric codes is provided in Appendix~\ref{sec:toric}.

\begin{figure}
	\centering 
	\subfloat[\label{fig:lattice_5x5}]{\includegraphics[height=0.20\textwidth]{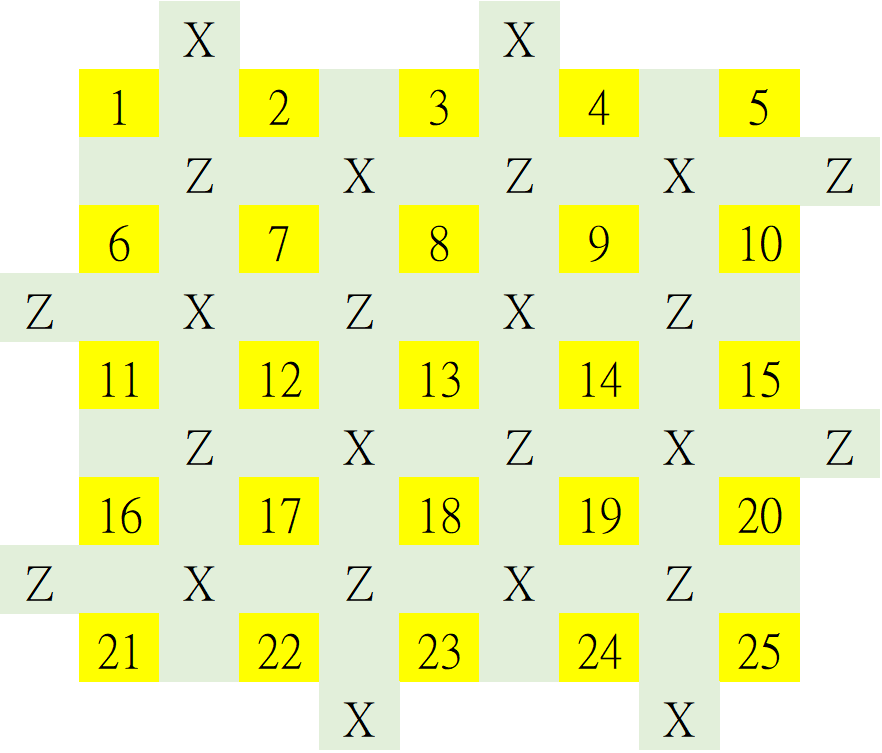}} ~~~~~~ 
	\subfloat[\label{fig:lattice_4x4}]{\includegraphics[height=0.18\textwidth]{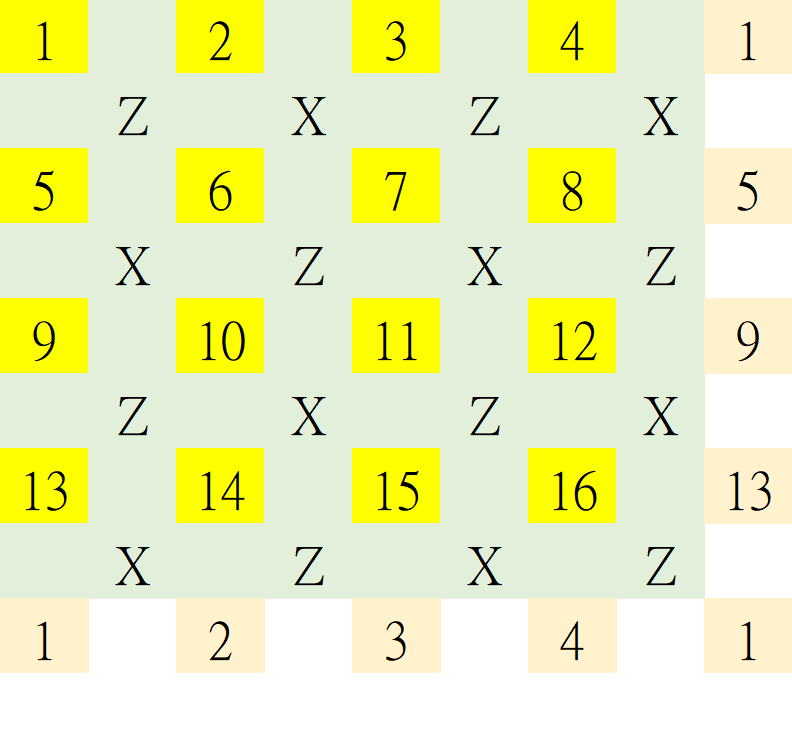}} 
	\caption{
		The lattice representations of (rotated) surface and toric codes.
		(a) $[[L^2, 1, L]]$ surface code with $L=5$. 
		(b) $[[L^2, 2, L]]$ toric code with $L=4$.
		In both figures, a~qubit is represented by a yellow box numbered from 1 to $N$. 
		 Since the toric code is defined on a torus, there are orange boxes on the right and bottom in (b), each representing the qubit of the same number. 
		An $X$- or $Z$-type stabilizer is indicated by a label $W\in\{X,Z\}$ between its neighboring qubits.
		For example, in~(a), the label $X$ between qubits ${1,2}$ is ${X_1 X_2}$
		and the label $Z$ between qubits ${1,2,6,7}$ is ${Z_1Z_2Z_6Z_7}$.
	} \label{fig:lattice}		
\end{figure}

An ${[[L^2,\, 1,\, L]]}$ surface code can be defined on an ${ L\times L }$ square lattice for $L\ge 3$.
Figure~\ref{fig:lattice}\,(a) shows an example of $L=5$. 
The stabilizer generators are of weight 2 or 4, independent of $L$. 
Thus, a large surface code has strong degeneracy, containing many stabilizers of weight $\ll D$.

When decoding surface codes, conventional BP often gets trapped. 
It is possible to use normalized BP to enlarge the step-size, but it has the side effect of causing BP to easily diverge.
MBP, with fixed inhibition, does not have this side effect and finds degenerate errors with high probability.
Two examples of $L=7$ are provided in Appendix~\ref{sec:surf2} to explain the improvement. 
The examples demonstrate how serial \ourBP exploits the degeneracy of surface codes.

\begin{figure}
	\centering \includegraphics[width=0.5\textwidth]{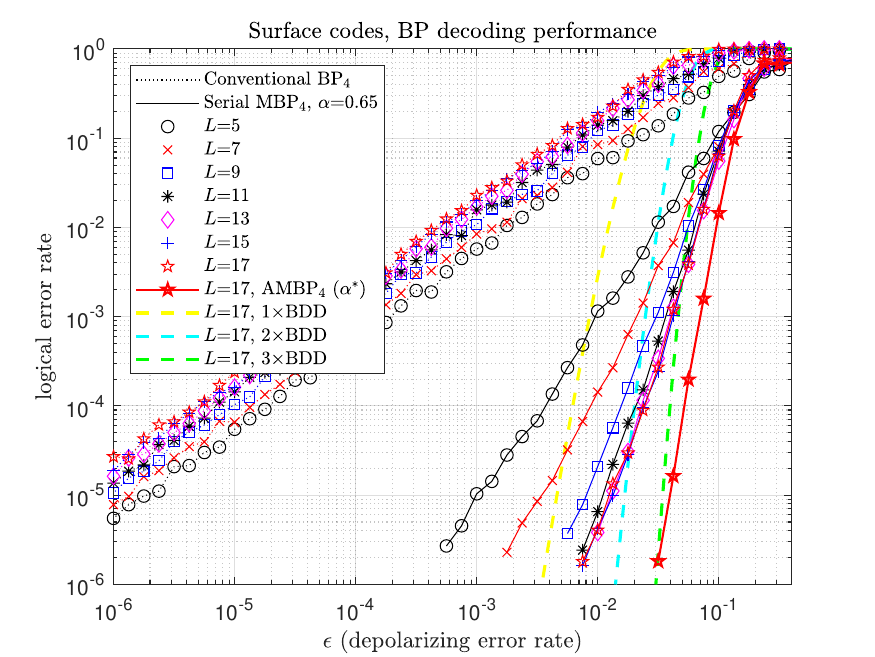}
	\caption{
		Performances of conventional BP$_4$ and serial MBP$_4$ on surface codes, based on $T_{\max}=150$.
		For generating this figure, we use a fixed $\epsilon_0 = 0.013$ to prevent the curve fluctuation 
		(as discussed in Remark~\ref{rmk:fix_eps_0}).
	} \label{fig:surf_65}		
\end{figure}

The performances of (M)BP$_4$ on surface codes are shown in Fig.~\ref{fig:surf_65}.
It can be seen that conventional BP$_4$ does not work on surface codes; even worse, the logical error rate is higher for larger distance. 
We remark that normalized BP$_4$ \eq{eq:gamma_nmN} is not effective either.
On the other hand, using serial MBP$_4$ with $\alpha<1$  significantly improves the performance.

\begin{figure}
	\centering \includegraphics[width=0.49\textwidth]{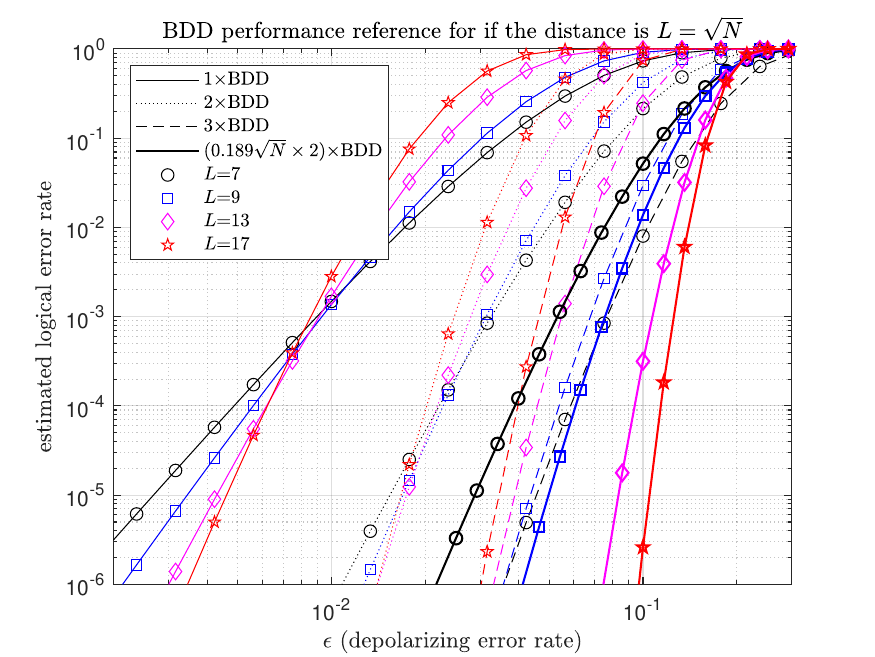}
	\caption{
		Performances of $r\times$BDD on codes of minimum distance $L$ and length $N=L^2$.
		The cases of $r=1,2,3$, and $0.189\sqrt{N}\times 2$ are plotted.
	} \label{fig:BDD_BLER}		
\end{figure}

Several BDD performance curves for $L=17$ are also provided in Fig.~\ref{fig:surf_65}. 
  Gallager expected   BP to have performance around $1\times$BDD to $2\times$BDD \cite{Gal63}.
Serial MBP$_4$ with $\alpha=0.65$ achieves this. However, this is not enough for surface codes.
In Fig.~\ref{fig:BDD_BLER}, we consider $r\times$BDD performances for $[[L^2,1,L]]$ codes at $r=1,2,3$.
The depolarizing rate at which the curves of the same $r$  intersect is called the ``$r\times$BDD error threshold" for the $[[L^2,1,L]]$ codes.
The theoretical error threshold for a two-dimensional code is 18.9\% \cite{WL12,BAOKM12,Ohz12}, and thus we would need a decoder with correction radius up to $0.189N$, 
which needs $r\times$BDD with non-fixed $r$ scaled as $0.189\sqrt{N}\times2$. 
In other words, as $L$ increases, an optimum decoder needs to correct most errors in a radius roughly equal to $0.189N$, despite that the  minimum distance only scales with $\sqrt{N}$.
The $(0.189\sqrt{N}\times2)\times$BDD   curves are also drawn in  Fig.~\ref{fig:BDD_BLER} (bold lines) and their intersection suggests an error threshold of roughly 18.9\%.

We observed that the performance of \ourBP on surface codes saturates for large $L$, i.e., the slope of the performance curve does not increase when $L$ gets larger.
(A similar phenomenon is observed in the neural BP \cite[Fig.~4(b)]{LP19}.)
Using  AMBP$_4$ with $\alpha^* \in \{1.0, 0.99, \dots, 0.5\}$, 
we have much improved performance for $L=17$ as shown in Fig.~\ref{fig:surf_65}, at the cost of higher computation complexity.

\begin{figure*}
	\subfloat[\label{fig:surf_dg}]{\includegraphics[width=0.34\textwidth]{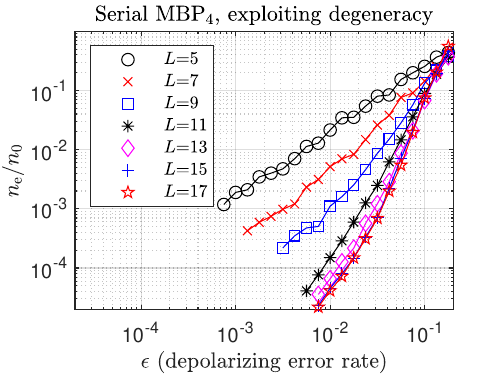}}  
	\subfloat[\label{fig:surf_fa}]{\includegraphics[width=0.34\textwidth]{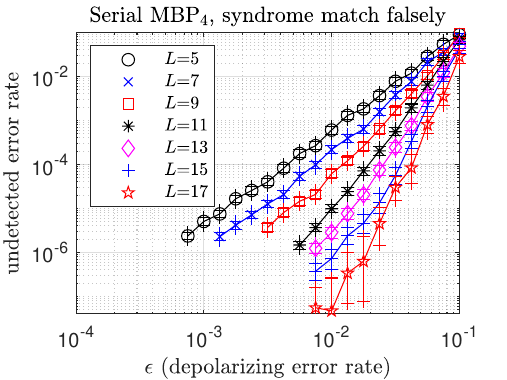}} 
	\subfloat[\label{fig:surf_it}]{\includegraphics[width=0.34\textwidth]{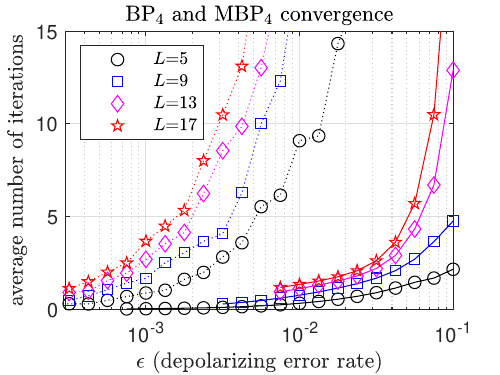}} 
	\caption{
		Some statistics of decoding surface codes using serial MBP$_4$ with $\alpha=0.65$ (solid lines).
		(a)~The ratio $n_{\rm e}/n_{\rm 0}$.
		(b)~Undetected error rate. 
		(c)~Average numbers of iterations; also shown in (c) are the numbers for conventional BP$_4$~(dotted~lines).
	} \label{fig:surf}
\end{figure*}

\begin{figure}
	\centering \includegraphics[width=0.5\textwidth]{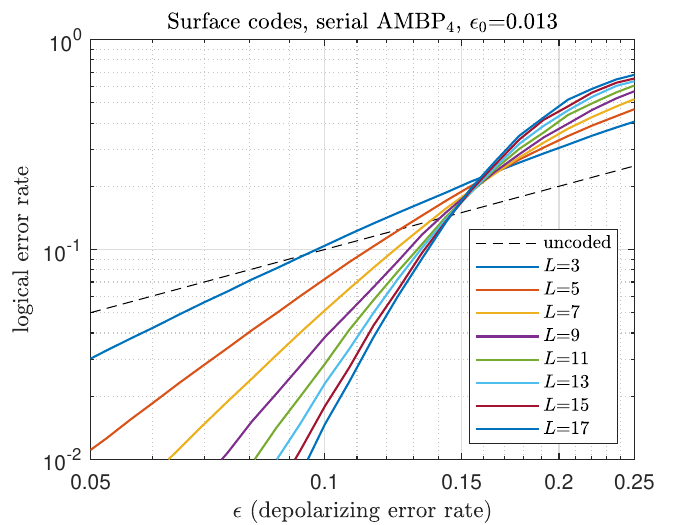}
	\caption{
		The threshold performance of serial AMBP$_4$ on surface codes, which is roughly 16\%.
		The dashed line stands for the case of no error correction (logical error rate $=\epsilon$).
	} \label{fig:thld_surf_ais}
\end{figure}

Note that  a maximum number of iterations $T_{\max}=150$ is allowed in the simulations so that serial MBP$_4$ with a fixed~$\alpha$ can perform better.
If proper $\alpha^*$ can be selected, the value of $T_{\max}$ can be reduced a lot (e.g., to 30) without significant performance loss.
In fact, sometimes it is possible to use a smaller value of $\alpha$ instead of the $\alpha^*$ above, and the convergence can be even faster and remain correct.
We discuss this possibility in Appendix~\ref{sec:surf2} (see \eq{eq:d7_ai50_it2}).

Next we explain how \ourBP has improvement over BP$_4$.   
Again consider serial \ourBP with $\alpha=0.65$. We examine the types of corrected errors as in \eq{eq:degen}.  
The ratio $n_{\rm e}/n_{\rm 0}$ becomes smaller if the decoder can exploit the degeneracy, which is indeed the case of \ourBP as shown in Fig.~\ref{fig:surf_dg}.
Although it is not shown, BP$_4$ has (bad) $n_{\rm e}/n_{\rm 0}\approx 1$, but (good) undetected error rate $\approx 0$ for $L>7$.
The  improvement of serial MBP$_4$ over BP$_4$ comes at a cost of non-negligible  undetected error rate, as shown in Fig.~\ref{fig:surf_fa}.
(A similar phenomenon is also observed in the neural BP \cite[Fig.~2(d)]{LP19}.)
Undetected error events occur because we use a larger step-size, so BP may jump far to a logical error with the syndrome falsely matched.
However, this is not a random search, or otherwise the ratio $n_{\rm e}/n_{\rm 0}$ would be as large as $1 - (1/2^{2K}) = 3/4$.
Also observe that, in Fig.~\ref{fig:surf_fa}, the undetected error rate becomes smaller as $L$ increases.

The average numbers of iterations are shown in Fig.~\ref{fig:surf_it}.
Serial MBP$_4$ spends much fewer iterations (which means much lower complexity), while BP$_4$ spends much more.
It means that MBP$_4$ improves from BP$_4$ by exhibiting better convergence behavior, 
rather than relying on increasing complexity through performing more iterations. 
Recall that better performance does not necessarily apply better convergence in average number of iterations (Fig.~\ref{fig:3786_it}).

Figure~\ref{fig:thld_surf_ais} is provided for observing the decoding threshold of AMBP$_4$ on surface codes, which is roughly 16\%.

The analysis of (A)MBP$_4$ on toric codes is similar and given in Appendix~\ref{sec:toric}. 
A threshold of roughly 17.5\% is observed for AMBP$_4$ on toric codes (Fig.~\ref{fig:thld_toric_ais}).

\pagebreak

\section{Conclusion and Future Research} \label{sec:Conclu}

We analyzed the energy topology of BP for decoding quantum codes, and proposed (A)MBP$_4$,
which efficiently explores the degeneracy and significantly improves the performance over conventional BP, especially for highly-degenerate codes. 
(A)MBP$_4$ has complexity almost linear in the code length.
To further improve (A)MBP$_4$, one may try to analyze the behavior of those uncorrected errors.

Considering BP as an RNN, our proposed scheme has adjustable edge weights (gradient decent step-size) scaled by $1/\alpha$ and fixed inhibition strength.
The network introduces proper memory effects, allowing BP to resist wrong beliefs or accelerate the search. 

By using AMBP$_4$, the achieved thresholds are roughly 16\% and 17.5\% on surface and toric codes, respectively, 
at the cost of complexity to choose $\alpha^*$.

A method to choose $\alpha^*$ at a lower cost would be desired.
To determine $\alpha^*$, the information of the error syndrome or channel statistics could be useful. 
For example, a syndrome vector of high weight may correspond to an error of high weight 
and a large step-size may be needed.

We note that $\alpha$ may be estimated by density evolution (DE) \cite{RSU01,CF02a,CDE+05,RU05} or extrinsic information transfer (EXIT) charts \cite{Bri01,AKB04,LS06}. Some other methods are mentioned in \cite{Bri01}. 
However, cycles in the Tanner graph may affect the accuracy of these methods \cite{CDE+05,LS06}.
With cycles, \cite{CDE+05} suggested to consider a method in \cite{CF02b}, which estimates the initial step-size and emphasizes its importance. 
What we did in Sec.~\ref{sec:GD} (and Fig.~\ref{fig:gmn_inv}) also serves for this purpose.

One may introduce parameters $\alpha_{mn,i}$ and $\beta_{mn,i}$ for each edge $(m,n)$ at each iteration $i$, but this would require us to develop a strategy of choosing these parameters.
In Algorithm~\ref{alg:LLR-BP4}, there are more useful edges (e.g., the dotted lines in Fig.~\ref{fig:S2x3_RNN_a}) not considered in BP-based neural networks \cite{Nac+18,LP19}. 
With these added edges, an MBP-based neural network may be considered. The energy function can be defined as in \eq{eq:energy} or any appropriate one. 
For training the edge weights, their initial values are important \cite{SMDH13}.
Our simulation results suggest that $(\alpha_{mn,i},\beta_{mn,i})$ can be initialized as $(\alpha, 0)$, with possible disturbance if needed.

The values of $(\alpha_{mn,i},\beta_{mn,i})$ have some derivable trend as in \eq{eq:parJ}--\eq{eq:gamma_update2}.
For example, in Algorithm~\ref{alg:LLR-BP4}, we replace $1/\alpha$  by $g_{mn}$ derived for each $(m,n)$ at each iteration by \eq{eq:gmn}
up to a constant such that $g_{mn}=1$ at the first iteration;
then the decoder performance improves (see \cite{KL22isit} arXiv version).

We remark that surface (or toric) codes can be decoded by MBP$_4$ with a good parallelism. 
For example, consider the $[[16,2,4]]$ toric code in Fig.~\ref{fig:lattice}\,(b). 
The qubits can be divided into four groups $\{1,2,5,6\}$, $\{3,4,7,8\}$, $\{9,10,13,14\}$, and $\{11,12,15,16\}$. 
Then the qubits in each group can be decoded in a serial order, while all groups can be run simultaneously. 
Keeping a group size of four, we have $N/4$ groups, which if run with parallelism have $O(N/(N/4))=O(1)$ decoding time.
We simulate this decoding order for surface codes and a threshold of roughly 15.5\% is observed.

Our scalar-based approach could be extended to the case of fault-tolerant circuits.
We have an initial study in the data and syndrome error model \cite{KCL21}.
Extending this to the fully fault-tolerant model is our ongoing work.

\section*{Acknowledgment}	
CYL was financially supported from the Young Scholar Fellowship Program by the Ministry of Science and Technology (MOST) in Taiwan, under Grant MOST109-2636-E-009-004.

\bibliographystyle{IEEEtran}	
\bibliography{References_v7}

\newpage

\appendices	

\section{Runtime of MBP$_4$} \label{sec:runtime}

\begin{figure}
	\centering \includegraphics[width=0.45\textwidth]{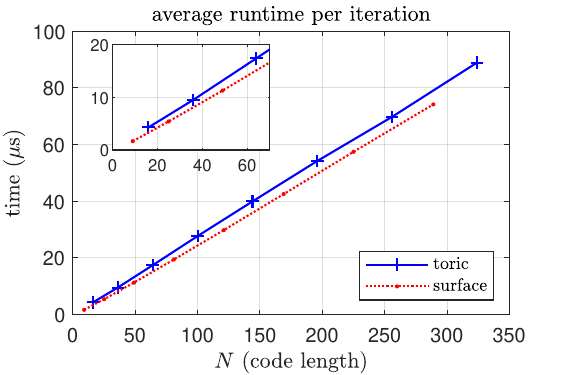}
	\caption{
		Almost linear runtime of MBP$_4$. The average runtime of MBP$_4$ on each toric or surface code is plotted.
	} \label{fig:runtime}
\end{figure}

We verify that the runtime of MBP$_4$ is $O(Nj)$  per iteration. First, consider toric codes, which have fixed column-weight ${j=4}$.
	We test serial MBP$_4$ with $\alpha=0.75$ at depolarizing rate $0.32$ on one core (4.9GHz) of an Intel i9-9900K machine.
	The average runtime per iteration is  shown in Fig.~\ref{fig:runtime},
	which  is obviously linear in $N$.
	Then we consider surface codes, which have  mean column-weight slightly smaller than $4$.
	As expected, the average runtime per iteration is again linear in $N$ and the slope is smaller than that for the toric codes, as shown in Fig.~\ref{fig:runtime}.

\section{More simulation results} \label{sec:MoreSim}

\begin{figure}
	\centering (a) \includegraphics[width=0.41\textwidth]{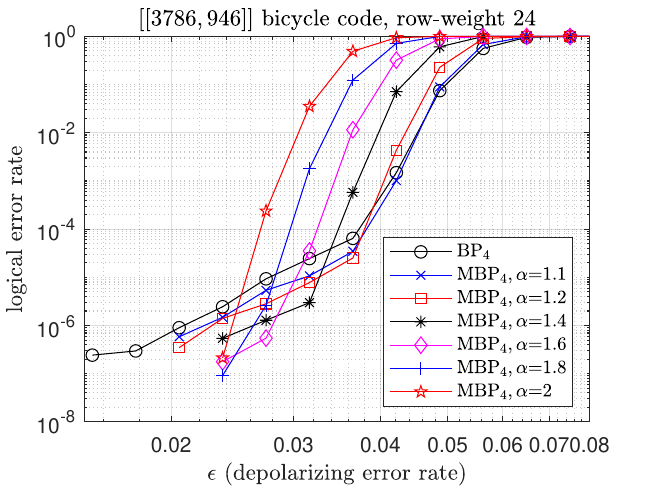}	\\ \vspace*{\floatsep}
	\centering (b) \includegraphics[width=0.41\textwidth]{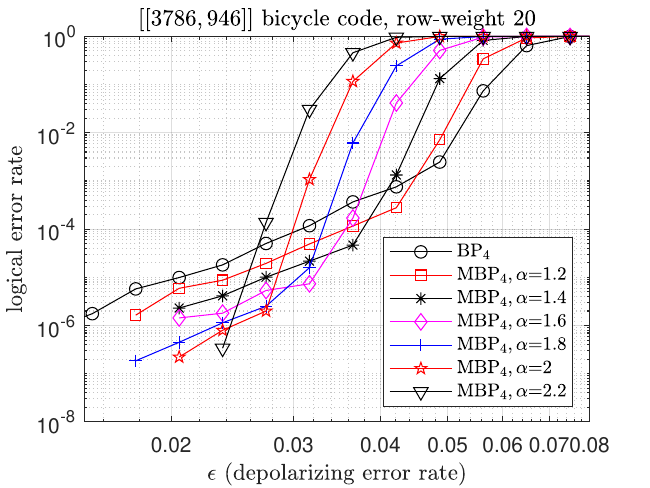}	\\ \vspace*{\floatsep}
	\centering (c) \includegraphics[width=0.41\textwidth]{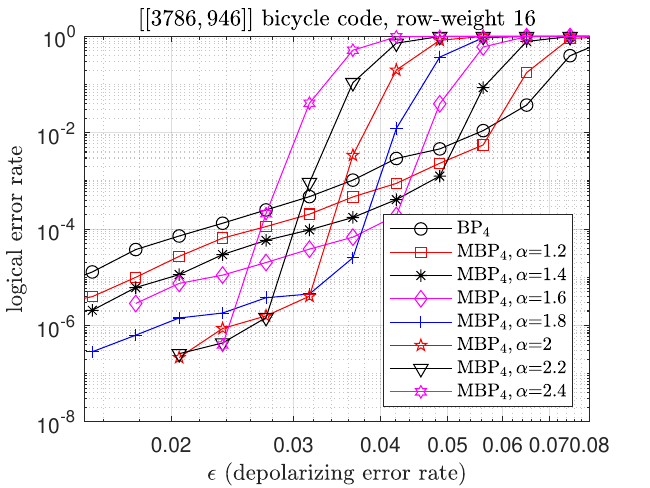}	\\ \vspace*{\floatsep}
	\centering (d) \includegraphics[width=0.41\textwidth]{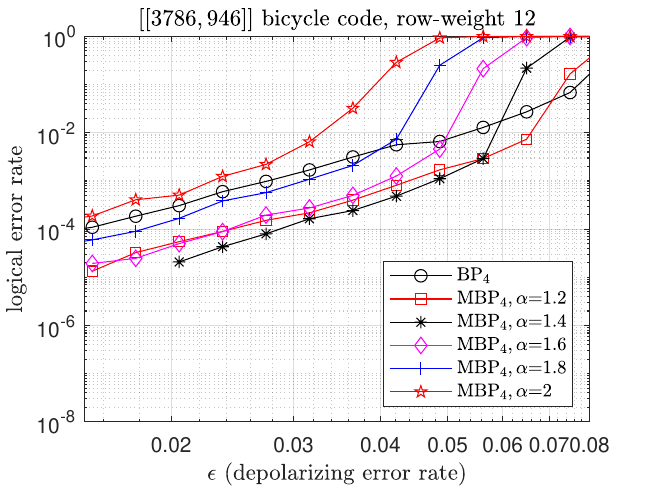}	
	\caption{
	Performances of parallel (M)BP$_4$ on the $[[3786,946]]$ bicycle codes with row-weights $k$. 
	\mbox{(a) $k=24$.}  \mbox{(b) $k=20$.}  \mbox{(c) $k=16$.}  \mbox{(d) $k=12$.}
	} \label{fig:3786_k}
\end{figure}

\subsection{Bicycle codes} \label{sec:bic2}

Herein we  provide more simulation results for the $[[3786,946]]$ bicycle codes with row-weight 24, 20, 16, or 12
in Figs.~\ref{fig:3786_k}\,(a)--(d), respectively. 
$T_{\max}=90$ is used.

The results match the expectations (Fig.~\ref{fig:gmn_inv}): a smaller $\epsilon$ needs a larger $\alpha$;
a larger $k$ needs a larger $\alpha$ before the performance saturation occurs at a smaller $\epsilon$.

A code of larger row-weight $k$ has a larger mean column-weight $kM/N$. 
Since bicycle tend to be nondegenerate, we recall some classical expectations \cite{Gal63,Mac99}:
a code of smaller column-weight has earlier curve rolling-off (at larger~$\epsilon$);
a code of larger row-weight has lower error-floor.
Our results match these expectations.

\subsection{Surface codes} \label{sec:surf2}

\begin{figure*}
	\begin{flushleft}
	~~~~~~ Physical error pattern 1 ~~~~~~~~~~~~~~~~~~~~~~~~~~~~~~~~~~ Parallel BP$_4$ (failure) ~~~~~~~~~~~~~~~~~~~~ Serial MBP$_4$ with $\alpha=0.65$ (success)\\
	~~~~~ ($X_4 Z_{15} Z_{16} Y_{23} Z_{33} Y_{39} Y_{40}$)
	~~~~~~~~~~~~~~~~~~~~~~~~~~~~~~~~~~~ ($X_{23} Z_{33} Y_{39} Y_{40}$)
	~~~~~~~~~~~~~~~~~~~~~~~~~~~~~~ ($X_3 Z_{22} X_{23} X_{32} Y_{33} Z_{39} Z_{40}$)
	\end{flushleft}
	\centering \includegraphics[width=1\textwidth]{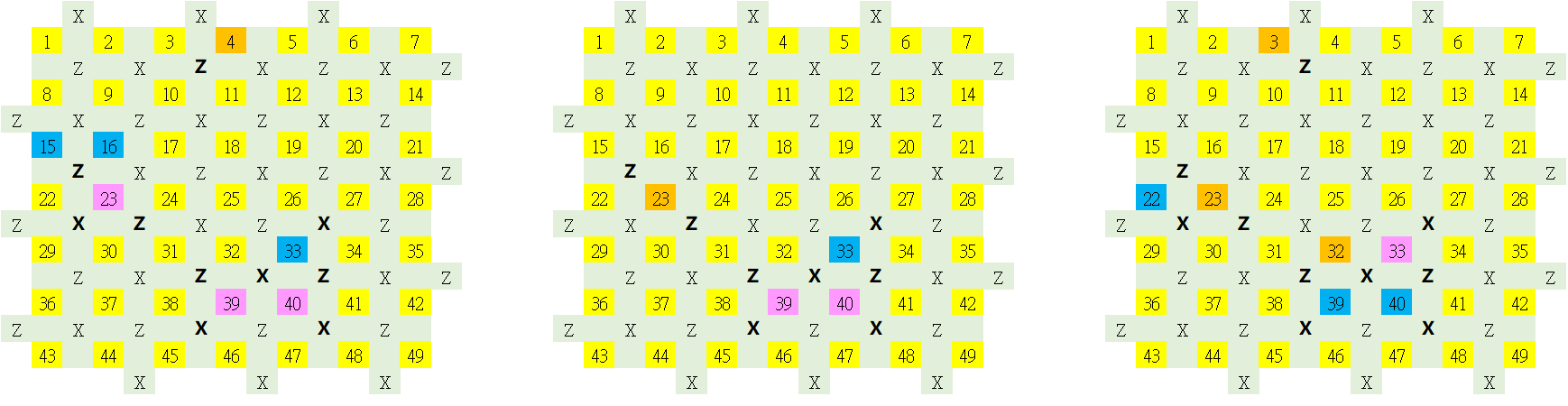} \\	
	(a)~~~~~~~~~~~~~~~~~~~~~~~~~~~~~~~~~~~~~~~~~~~~~~~~~~~~~~~~~~~~~~~~(b)~~~~~~~~~~~~~~~~~~~~~~~~~~~~~~~~~~~~~~~~~~~~~~~~~~~~~~~~~~~~~~~~(c)\\ \vspace*{\floatsep}
	\begin{flushleft}
	~~~~~ Physical error pattern 2 ~~~~~~~~~~~~~~~~~~~~~~~~~~~~~~~~~~~ Parallel BP$_4$ (failure) ~~~~~~~~~~~~~~~~~~~~ Serial MBP$_4$ with $\alpha=0.65$ (success)\\
	~~~ ($X_4 X_6 X_7 Z_{15} Z_{16} Y_{23} Z_{33} Y_{39} Y_{40}$)
	~~~~~~~~~~~~~~~~~~~~~~~~~~~~ ($X_7 X_{23} Z_{33} Y_{39} Y_{40}$)
	~~~~~~~~~~~~~~~~~~~ ($X_3 X_5 X_7 Z_{22} X_{23} X_{26} X_{27} Y_{33} X_{34} Y_{39} Y_{40}$)
	\end{flushleft}
	\centering \includegraphics[width=1\textwidth]{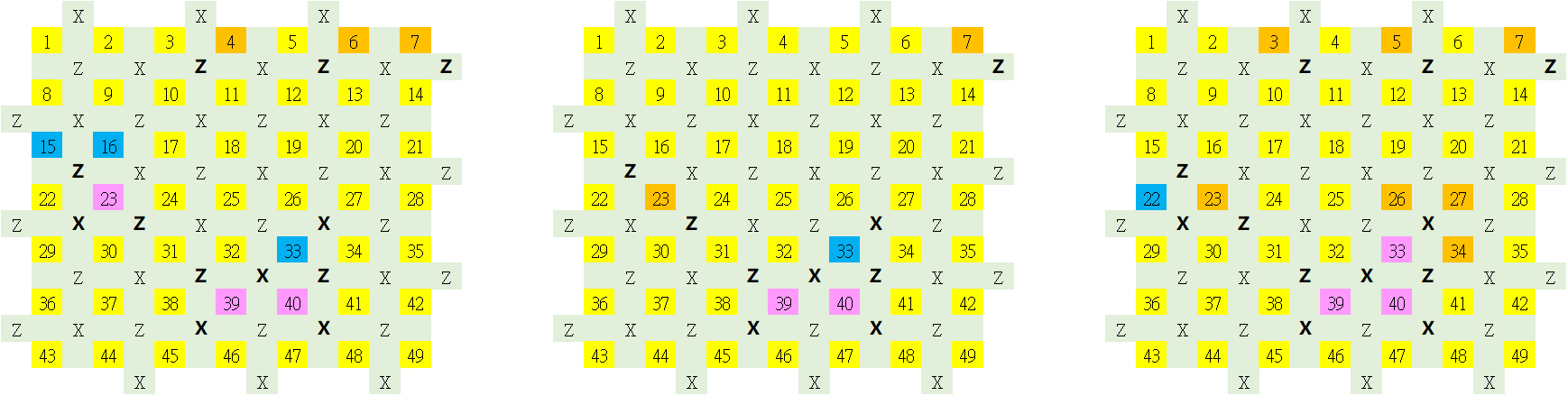} \\	
	(d)~~~~~~~~~~~~~~~~~~~~~~~~~~~~~~~~~~~~~~~~~~~~~~~~~~~~~~~~~~~~~~~~(e)~~~~~~~~~~~~~~~~~~~~~~~~~~~~~~~~~~~~~~~~~~~~~~~~~~~~~~~~~~~~~~~~(f)
	\caption{
		Two examples  of the $L=7$ surface code: (a)--(c) the first example and (d)--(f) the second example.
		A clean qubit is denoted by a numbered yellow box.  If a qubit suffers  an $X$, $Y$, or $Z$ error, it is denoted by an orange,  purple, or blue box, respectively. 
		A stabilizer is denoted by an $X$ or~$Z$. If the corresponding syndrome bit is 1, it is bold-faced. 
		Subfigures (a) and (d) are the actual error patterns, 
		(b) and (e) are the corresponding decoding results of conventional BP$_4$ (which are failures with unmatched syndromes), and 
		(c) and (f) are the corresponding decoding results of serial MBP$_4$ 
			(which are degenerate errors of those in (a) and (d), respectively).
	} \label{fig:surf_d7}
\end{figure*}

We provide two examples of the decoding on $L=7$ surface code in Figs.~\ref{fig:surf_d7}\,(a)--(c) and Figs.~\ref{fig:surf_d7}\,(d)--(f), respectively.

\begin{table*}
	\caption{
		The estimated $\hat E$ for decoding the error in Fig.~\ref{fig:surf_d7}\,(a) at each iteration (iter.) using the decoders in Fig.~\ref{fig:surf_d7}\,(b) and Fig.~\ref{fig:surf_d7}\,(c), as well as various decoder configurations in between. 
	} \label{tbl:d7_degen1} \centering
  \resizebox{1.02\textwidth}{!}{ 
	$\begin{array}{|c|l|l|l|}
	\hline 
	\text{iter.} & \text{Fig.~\ref{fig:surf_d7}\,(b) Parallel BP$_4$} & \text{Parallel normalized BP$_4$ with $\alpha_c=0.65$}   & \text{Parallel MBP$_4$ with $\alpha=0.65$}  \\ 
	\hline
	1	& Y_{23}X_{31}Y_{33}Y_{39}Y_{40} & X_{3} X_{4} Z_{22}Y_{23}Z_{29}X_{31}Y_{33}Y_{39}Y_{40}                                                                   & X_{3} X_{4} Z_{22}Y_{23}Z_{29}X_{31}Y_{33}Y_{39}Y_{40} \\
	2	& X_{23}Y_{33}Y_{39}Y_{40}       & X_{3} X_{4} X_{10}Z_{22}X_{23}Z_{24}Y_{29}Z_{30}Y_{31}X_{32}Y_{33}Y_{39}Y_{40}                                           & X_{23}Y_{29}Y_{33}Y_{39}Y_{40}                         \\
	3	& X_{23}Z_{33}Y_{39}Y_{40}		 & X_{3} X_{4} X_{16}Y_{22}Y_{24}X_{25}Z_{29}Y_{30}Z_{33}Y_{39}Z_{40}                                                       & X_{3} X_{4} Y_{22}X_{23}Z_{29}Z_{33}Y_{39}Z_{40}       \\
	4	& Y_{22}X_{23}Z_{33}Y_{39}Y_{40} & X_{3} X_{4} Y_{8} Z_{9} Y_{10}Z_{15}X_{16}Z_{18}Z_{25}Z_{31}Z_{32}Z_{33}Z_{39}Y_{40}                                     & X_{10}X_{23}Z_{33}Z_{39}Y_{40}                         \\
	5	& X_{23}Z_{33}Y_{39}Y_{40}       & Y_{9} X_{10}X_{16}Z_{26}Z_{30}X_{31}Y_{33}Z_{38}Y_{39}Y_{40}                                                             & Z_{22}X_{23}Z_{29}Z_{33}Y_{39}Y_{40}                   \\
	6	& X_{23}Z_{33}Y_{39}Y_{40}       & X_{4} X_{10}Y_{16}Z_{18}X_{27}Z_{33}Y_{39}Z_{40}Y_{44}X_{48}X_{49}                                                       & X_{3} X_{4} X_{23}Z_{33}Y_{39}Y_{40}                   \\
	7	& X_{23}Z_{33}Y_{39}Y_{40}       & X_{3} X_{4} X_{10}Y_{17}Z_{22}Y_{23}Y_{24}Z_{29}Y_{31}Z_{39}X_{40}                                                       & Z_{22}X_{23}Z_{29}Z_{33}Y_{39}Y_{40}                   \\
	8	& X_{23}Z_{33}Y_{39}Y_{40}       & X_{17}Y_{23}X_{29}Y_{31}Y_{33}Z_{34}Z_{39}Y_{40}                                                                         & X_{3} X_{4} X_{23}Z_{33}Y_{39}Y_{40}                   \\
	9	& X_{23}Z_{33}Y_{39}Y_{40}       & X_{3} X_{4} X_{10}Z_{16}Z_{21}X_{22}X_{23}Y_{24}Y_{26}X_{27}Z_{28}Y_{30}X_{32}Z_{33}Y_{38}Y_{39}Z_{40}Z_{41}Y_{46}X_{47} & Z_{22}X_{23}Z_{29}Z_{33}Y_{39}Y_{40}                   \\
	10& X_{23}Z_{33}Y_{39}Y_{40}       & Z_{1} Z_{4} X_{10}Y_{16}Y_{25}Y_{26}X_{27}Z_{30}Z_{31}Y_{32}X_{33}X_{34}X_{37}X_{40}X_{42}Y_{44}Y_{46}Y_{47}             & X_{3} X_{4} X_{23}Z_{33}Y_{39}Y_{40}                   \\
	11& \text{... (Get trapped)}       & \text{... (Diverge)}      & \text{... (Oscillate)} \\
	\hline
	\hline 
	\text{iter.} & \text{Serial BP$_4$} & \text{Serial normalized BP$_4$ with $\alpha_c=0.65$ } & \text{Fig.~\ref{fig:surf_d7}\,(c) Serial MBP$_4$ with $\alpha=0.65$  } \\ 
	\hline
	1	& X_{23}Y_{33}                         & X_{3}X_{4} X_{5} X_{6} X_{7} Z_{22}Y_{23}Z_{29}X_{31}Y_{33}Y_{39}Y_{40}                                                       & X_{3} Z_{22}X_{23}Y_{33}Y_{40}                                           \\
	2	& X_{23}X_{31}X_{32}Y_{33}Z_{38}Y_{39} & X_{3}X_{4} X_{5} X_{6} X_{7} X_{10}X_{11}X_{12}X_{13}X_{14}Z_{22}X_{23}Z_{24}Y_{29}Z_{30}Y_{31}X_{32}Y_{33}Y_{39}Y_{40}       & X_{3} X_{23}Z_{29}Z_{30}X_{32}Y_{33}Z_{38}Z_{39}Z_{40}                   \\
	3	& Y_{29}Y_{30}Z_{33}Y_{39}Y_{40}       & X_{3}X_{4} X_{7} X_{11}X_{12}X_{13}X_{14}X_{16}Y_{22}Y_{24}X_{25}Z_{29}Y_{30}Z_{33}Y_{39}Z_{40}                               & X_{23}Z_{30}Y_{33}Y_{39}Y_{40}                                           \\
	4	& Y_{22}Z_{33}Y_{39}Y_{40}             & Y_{8}Z_{9} Y_{10}Y_{11}X_{13}X_{14}Z_{15}X_{16}Z_{18}Z_{25}Z_{31}Z_{32}Z_{33}Z_{39}Y_{40}                                     & X_{3} X_{23}X_{32}Z_{33}Z_{39}Z_{40}                                     \\
	5	& X_{23}Z_{33}Y_{39}Y_{40}             & Y_{9}Y_{12}X_{16}X_{23}Z_{26}Z_{30}X_{31}Y_{33}Z_{38}Y_{39}Y_{40}                                                             & X_{3} Z_{22}X_{23}Z_{33}X_{34}Y_{39}Y_{40}                               \\
	6	& X_{23}Z_{33}Y_{39}Y_{40}             & X_{4}X_{7} Z_{13}Y_{16}Z_{18}Z_{19}Y_{20}X_{27}Y_{28}Z_{30}Z_{33}Y_{39}Z_{40}Y_{44}X_{48}X_{49}                               & X_{3} Z_{22}X_{23}X_{25}X_{26}X_{27}X_{32}Y_{33}X_{34}Z_{39}Z_{40}X_{46} \\
	7	& X_{23}Z_{33}Y_{39}Y_{40}             & X_{3}X_{4} Y_{7} X_{10}Y_{13}Y_{17}X_{21}Y_{23}Y_{24}Y_{31}Z_{39}X_{40}                                                       & X_{3} X_{17}Z_{22}X_{23}X_{24}X_{26}X_{27}X_{31}Z_{33}Z_{39}Z_{40}       \\
	8	& X_{23}Z_{33}Y_{39}Y_{40}             & X_{7}Y_{10}X_{17}Y_{23}Z_{27}Y_{31}Z_{32}Z_{33}Z_{34}Y_{39}Y_{40}                                                             & X_{3} X_{10}Z_{22}X_{23}X_{26}X_{32}Y_{33}Z_{39}Z_{40}                   \\
	9	& X_{23}Z_{33}Y_{39}Y_{40}             & X_{3}X_{4} X_{5} X_{6} X_{7} X_{10}X_{14}Z_{21}X_{23}Z_{24}Y_{26}Z_{28}Y_{33}Y_{34}Z_{38}Y_{39}Y_{40}Z_{41}Y_{46}X_{47}       & X_{11}Z_{22}X_{23}X_{32}Y_{33}Z_{39}Z_{40}                               \\
	10& X_{23}Z_{33}Y_{39}Y_{40}             & Z_{1}Z_{4} X_{5} X_{6} X_{12}Y_{13}X_{14}Y_{16}Y_{18}Y_{21}X_{23}Y_{25}X_{26}X_{27}Y_{28}Z_{30}Z_{41}Y_{42}Y_{47}X_{48}X_{49} & X_{3} Z_{22}X_{23}X_{32}Y_{33}Z_{39}Z_{40}~~\text{(a degenerate error)}                              \\
	11& \text{... (Get trapped)}             & \text{... (Diverge)}      & \\ 
	\hline
	\end{array}$
  }		
\end{table*}

The first example is an error of weight 7 as in Fig.~\ref{fig:surf_d7}\,(a). 
The error $X$ at qubit 4 anticommutes with stabilizer $Z_3Z_4Z_{10}Z_{11}$, so the corresponding syndrome bit is 1.
Conventional BP$_4$ (parallel BP$_4$) cannot decide whether an error $X$ is at qubit 3 or qubit 4.
To break this symmetry, possible methods  include post-processes (random perturbation \cite{PC08}, 
OSD \cite{PK19,RWBC20}, etc) or pre-processes (like training \cite{TM17,KJ17,LP19,MKJ19}). 
On the other hand, using serial MBP$_4$ with $\alpha=0.65$ can quickly decide a degenerate error $X$ at qubit~3, without additional processes.

Now we consider the complete error set on the surface code Fig.~\ref{fig:surf_d7}\,(a). 
The update of the estimated error at each iteration is shown in Table~\ref{tbl:d7_degen1} for various combinations of BP decoding.
	Conventional BP$_4$ (no matter parallel or serial) gets trapped in the same small-weight error Fig.~\ref{fig:surf_d7}\,(b).
	Normalized BP$_4$ with $\alpha=0.65$ (no matter parallel or serial) diverges.
	Parallel MBP$_4$ with $\alpha=0.65$ can approach the solution but oscillates. 
	Serial MBP$_4$ with $\alpha=0.65$ effectively converges to a degenerate error Fig.~\ref{fig:surf_d7}\,(c).

The reason that BP$_4$ easily gets trapped around a small-weight error is explained in Sec.~\ref{sec:J_S}. 
Serial MBP$_4$ with $\alpha=0.65$ performs a fast asymmetric update with large step-size and fixed inhibition (with provides memory effects for convergence, cf.~Fig.~\ref{fig:chk_Y4_a15}). 
As shown in Table~\ref{tbl:d7_degen1}, it performs an aggressive search without jumping wrongly, and converges to a degenerate error Fig.~\ref{fig:surf_d7}\,(c), equivalent to the actual error Fig.~\ref{fig:surf_d7}\,(a) up to three stabilizers 
$X_3X_4$, $Z_{15}Z_{16}Z_{22}Z_{23}$, and $X_{32}X_{33}X_{39}X_{40}$.

In the second example, two more $X$ errors at qubits 6 and 7 are added. Now the overall error is as in Fig.~\ref{fig:surf_d7}\,(d) and of weight $9>D=7$. 
Similar to the previous example, conventional BP$_4$ gets trapped around a small-weight error Fig.~\ref{fig:surf_d7}\,(e).
Serial MBP$_4$ with $\alpha=0.65$ finds a degenerate error (of weight 11) in Fig.~\ref{fig:surf_d7}\,(f), 
equivalent to the actual error in Fig.~\ref{fig:surf_d7}\,(d) up to four stabilizers 
$X_3X_4$, $X_5X_6$, $Z_{15}Z_{16}Z_{22}Z_{23}$, and $X_{26}X_{27}X_{33}X_{34}$.

To demonstrate of the effects of a further smaller $\alpha$,  we reduce $\alpha$ from 0.65 to 0.5 in the two examples. 
In each case, the decoding correctly converges in two iterations:
	\begin{equation} \label{eq:d7_ai50_it2} 
	\begin{aligned}
	&\text{Serial MBP$_4$ with $\alpha=0.5$ on the error in Fig.~\ref{fig:surf_d7}\,(a)}:\\ 
	&\text{~~~~~~~~} \to X_3 Z_{22} X_{23} Y_{33} X_{39} Y_{40}\\
	&\text{~~~~~~~~} \to X_3 X_{23} Z_{29} X_{32} Y_{33} Z_{39} Z_{40}.\\
	&\text{Serial MBP$_4$ with $\alpha=0.5$ on the error in Fig.~\ref{fig:surf_d7}\,(d)}:\\ 
	&\text{~~~~~~~~} \to X_{3} X_{5} X_{7} Z_{22} X_{23} Y_{33} X_{39} Y_{40}\\
	&\text{~~~~~~~~} \to X_{3} X_{5} X_{7} X_{23} Z_{29} X_{32} Y_{33} Z_{39} Z_{40}.
	\end{aligned}
	\end{equation}
The first result in \eq{eq:d7_ai50_it2} is equivalent to the error in Fig.~\ref{fig:surf_d7}\,(a) up to four stabilizers
$X_3X_4$,\, $Z_{15}Z_{16}Z_{22}Z_{23}$,\, $Z_{22}Z_{29}$, and $X_{32}X_{33}X_{39}X_{40}$.
The second result in \eq{eq:d7_ai50_it2} is equivalent to the error in Fig.~\ref{fig:surf_d7}\,(d) up to five stabilizers
$X_3X_4$,\, $X_5X_6$,\, $Z_{15}Z_{16}Z_{22}Z_{23}$,\, $Z_{22}Z_{29}$, and $X_{32}X_{33}X_{39}X_{40}$.
This suggests that using a smaller $\alpha$ may have a faster and yet correct convergence.

Note that, as shown in Table~\ref{tbl:d7_degen1}, using normalized BP$_4$ with $\alpha_c<1$ makes the decoding to diverge.
This can also be observed by monitoring the variation of the  energy function and  we will discuss it in Appendix~\ref{sec:eta}.

\subsection{Toric codes} \label{sec:toric}

Now we give the decoding performance for $[[L^2,2,L]]$ toric codes with even $L$. An example of the toric code lattice is shown in Fig.~\ref{fig:lattice}\,(b).
		Note that  a toric code has every stabilizer generator associated with four qubits, and every qubit involved in four stabilizer measurements.
Thus the corresponding Tanner graph is regular, and every qubit is equally-protected.
This makes the toric code more suitable to have the same $\alpha$ for all the edges.

\begin{figure}[b]
	\centering \includegraphics[width=0.5\textwidth]{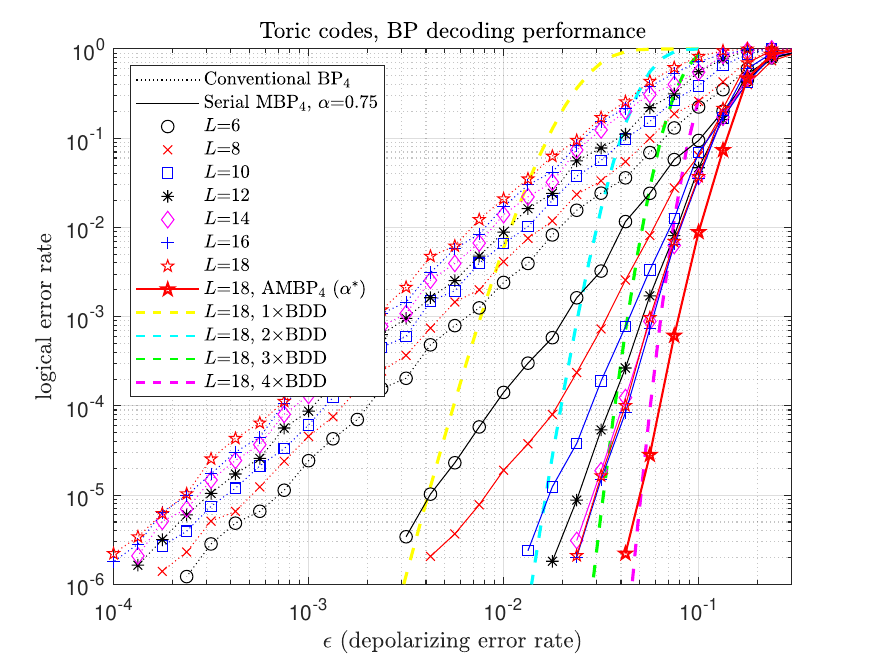}
	\caption{
		Performances of conventional BP$_4$ and serial MBP$_4$ on toric codes, based on $T_{\max}=150$.
		The curve of AMBP$_4$ ($\alpha^*$) is generated by fixed initialization $\epsilon_0 = 0.001$.
	} \label{fig:toric_75}		
\end{figure}

Figure~\ref{fig:toric_75} provides the simulation results of the toric codes with various distances. 
Since every qubit is equally-protected,
the performance of each toric code is generally better than the surface code of comparable size in Fig.~\ref{fig:surf_65}.
Note that  the performance curve of BP has no fluctuation with or without fixed initialization.

\begin{figure*}
	\subfloat[\label{fig:toric_dg}]{\includegraphics[width=0.34\textwidth]{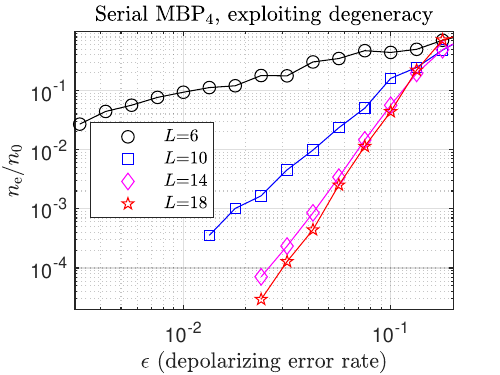}}  
	\subfloat[\label{fig:toric_fa}]{\includegraphics[width=0.34\textwidth]{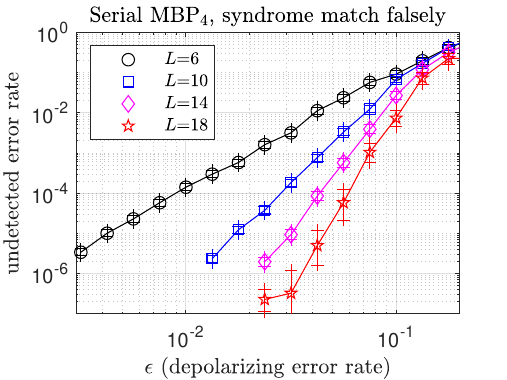}} 
	\subfloat[\label{fig:toric_it}]{\includegraphics[width=0.34\textwidth]{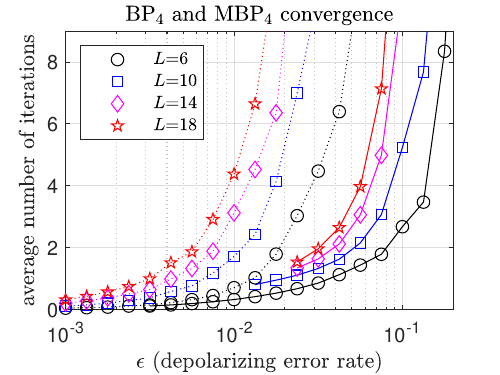}} 
	\caption{
		Some statistics of decoding toric codes using serial MBP$_4$ with $\alpha=0.75$ (solid lines).
		(a)~The ratio $n_{\rm e}/n_{\rm 0}$.
		(b)~Undetected error rate. 
		(c)~Average numbers of iterations; also shown in (c) are the numbers for conventional BP$_4$ (dotted lines).
	} \label{fig:toric}
\end{figure*}

\begin{figure}
	\centering \includegraphics[width=0.5\textwidth]{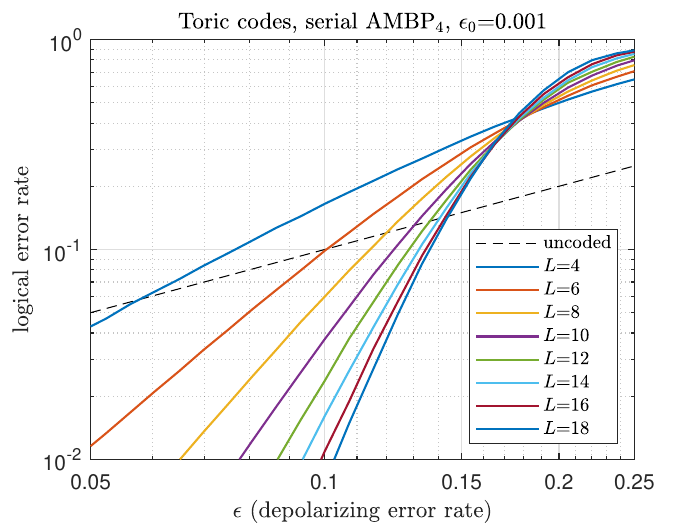}
	\caption{
		The threshold performance of serial AMBP$_4$ on toric codes, which is roughly 17.5\%. 
		The dashed line stands for the case of no error correction (logical error rate $=\epsilon$).
	} \label{fig:thld_toric_ais}			
\end{figure}

Our simulations suggest to use $\alpha=0.75$ in \ourBP for better performance, which is larger than $\alpha = 0.65$ used for surface codes (Fig.~\ref{fig:surf_65}).
This agrees with the observation in Fig.~\ref{fig:gmn_inv} that a code with larger row-weight should choose larger $\alpha$, 
since a toric code has fixed row-weight $4$, which is larger than the mean row-weight (between 2 and 4) of a surface code.

Similar to Fig.~\ref{fig:surf_65}, the performance of MBP$_4$ saturates when $L$ gets larger.  
 This can be improved by  AMBP$_4$ with  $\alpha^* \in \{1.0, 0.99, \dots, 0.5\}$.
We show a case ``$L=18$,~AMBP$_4$'' and several BDD curves in Fig.~\ref{fig:toric_75}.
As can be seen, AMBP$_4$ can correct most errors within $4\times$BDD correction radius in this case.
This is better than a comparable case ``$L$=17,~AMBP$_4$'' in Fig.~\ref{fig:surf_65}, which achieves roughly $3\times$BDD. 

By initializing $\Lambda$ with respect to a fixed $\epsilon_0=0.001$, 
we get a slightly better interpolation performance for large $\epsilon$ so we apply this technique when choosing $\alpha^*$.

Similar to Fig.~\ref{fig:surf}, we provide some statistics for decoding toric codes in Fig.~\ref{fig:toric}.
Again, serial MBP$_4$ with $\alpha<1$ significantly improves the conventional BP$_4$ by exploiting degeneracy (Fig.~\ref{fig:toric_dg}) at the cost of having some undetected errors (Fig.~\ref{fig:toric_fa}), and  MBP$_4$ has better algorithm convergence (Fig.~\ref{fig:toric_it}).

By using serial AMBP$_4$ on toric codes, we observe a threshold of roughly 17.5\%, as shown in Fig.~\ref{fig:thld_toric_ais}.

\section{Energy function}\label{sec:eta}

\begin{figure*}
	\subfloat[\label{fig:Ex1_f_B}]  {\includegraphics[width=0.52\textwidth]{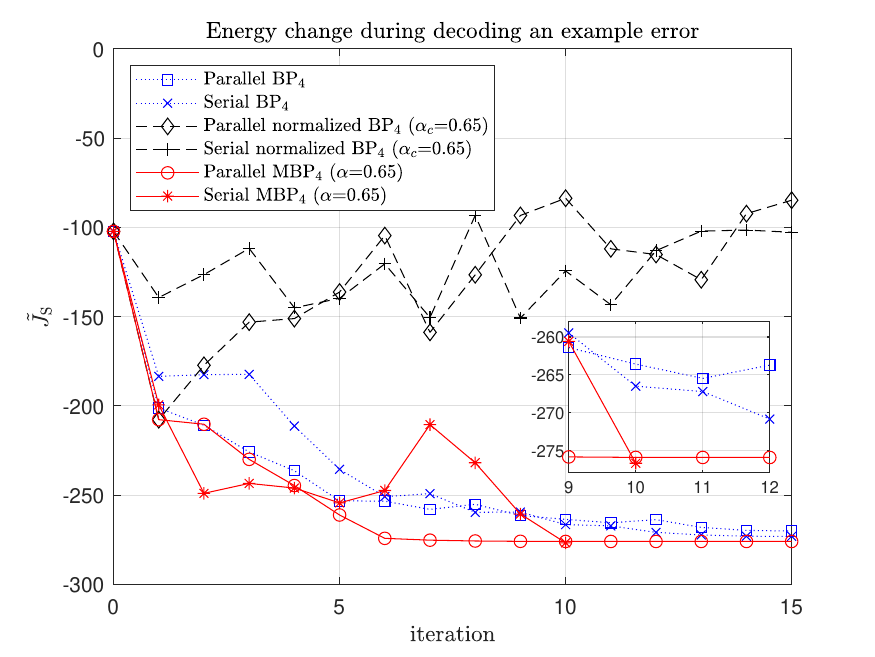}} 
	\subfloat[\label{fig:Ex1_f_dot}]{\includegraphics[width=0.52\textwidth]{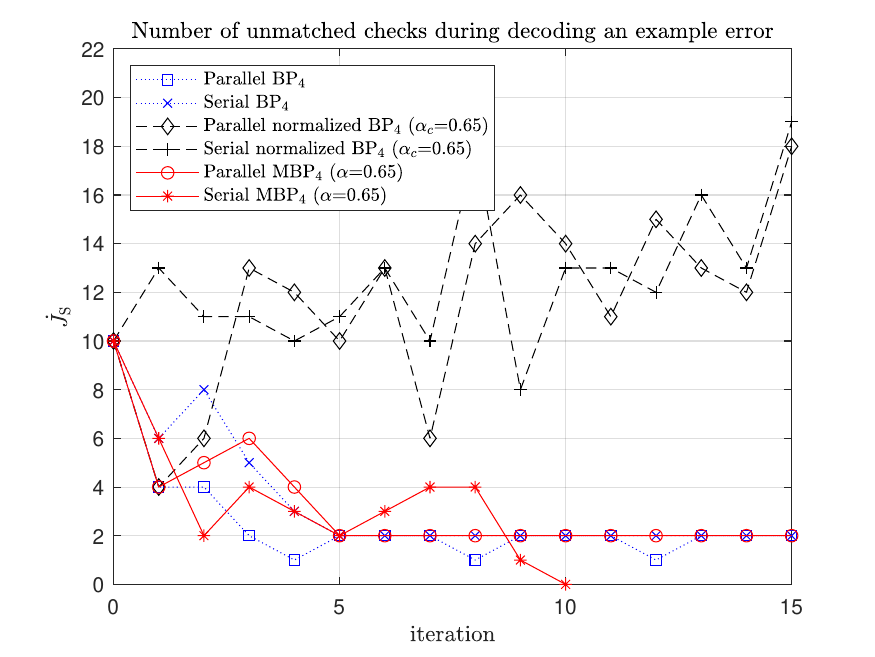}}
	\caption{
		Plotting the change of the energy function during iterations, for decoding the error pattern in Fig.~\ref{fig:surf_d7}\,(a) with the six configurations in Table~\ref{tbl:d7_degen1}.
		(a) Using $\tilde{J}_\text{S}$ in \eq{eq:tilde_f} with $B=6$.
		(b) Using $\dot{J}_\text{S}$ in \eq{eq:dot_f}.
		Serial MBP$_4$ with $\alpha=0.65$ is the only configuration that achieves a successful decoding. 
	} \label{fig:Ex1_f}				
\end{figure*}

	We analyze the function $J_\text{S}$ in \eq{eq:energy2}. Since it is unbounded,
	we may consider a bounded $\tilde{J}_\text{S}$ from $J_\text{S}$ by 
	\begin{equation} \label{eq:tilde_f}
	\tilde{J}_\text{S} = -\sum_{m=1}^M \tilde{\Delta}_m ~\approx~ -\sum_{m=1}^M \Delta_m,  
	\end{equation}
	where $\Delta_m = 2   \tanh^{-1}\left( (-1)^{z_m} \prod_{n\in \cN(m)} \tanh\left(  \frac{\lambda_{S_{mn}}( \Gamma_n)}{2}   \right) \right)$
	and $\tilde{\Delta}_m \approx \Delta_m$ but bounded as
	\begin{equation} \label{eq:Dm_bound_by_B}
	\tilde{\Delta}_m = \sgn(\Delta_m)\min\{|\Delta_m|, B\}
	\end{equation}
	for some positive $B$. 

	Note that if we consider MBP$_4$ in linear domain (see Algorithm~\ref{alg:QBP_M}) and define
	$$\delta_m = \textstyle \prod_{n\in\sN(m)} \left( (q_n^{I} + q_n^{S_{mn}}) - (\sum_{W\in\{X,Y,Z\}\setminus S_{mn}} q_n^{W}) \right)$$ 
	according to the output distributions $\{(q_n^I,q_n^X,q_n^Y,q_n^Z)\}_{n=1}^N$, then  $\Delta_m = \ln\frac{1+\delta_m}{1-\delta_m}$ and 
	$B=6$ can cover $|\delta_m| \le 0.99$ since $\ln\frac{1+0.99}{1-0.99}\approx 5.3 < 6$.
	
	To have a differentiable function, we consider a Taylor expansion of $\Delta_m = \ln\frac{1+\delta_m}{1-\delta_m}$ at $\delta_m=0$ as
	\begin{equation} \label{eq:Dm_taylor}
	\tilde{\Delta}_m^{(\ell)}= 2\left( \delta_m+\frac{\delta_m^3}{3}+\frac{\delta_m^5}{5}+\dots+\frac{\delta_m^\ell}{\ell} \right)
	\end{equation}
	with a finite order $\ell\ge 1$. Let the corresponding approximation of $J_\text{S}$ be $\tilde{J}_\ell$, defined by 
	\begin{equation} \label{eq:f_taylor}
	\tilde{J}_\ell = -\sum_{m=1}^M \tilde{\Delta}_m^{(\ell)} ~=\, -2\sum_{m=1}^M\left( \delta_m+\frac{\delta_m^3}{3}+\frac{\delta_m^5}{5}+\dots+\frac{\delta_m^\ell}{\ell} \right).
	\end{equation}
	In general, a good approximation of $J_\text{S}$ requires $\ell$ to be large. 
	However, even with $\ell=1$, the function
	\begin{equation} \label{eq:f_taylor_1}
	\tilde{J}_1 = -2\sum_{m=1}^M \delta_m,
	\end{equation}
	up to a scalar of 2, 
	is sufficient to be an energy function for a classical syndrome decoding based on hard-decision (see \cite{BB89}, especially Lemma~2 and (21) therein).

	Since a target $\delta_m>0$ (a target $\Delta_m>0$),
	we may also consider another energy function to focus on those $\Delta_m$ not yet positive (i.e., to focus on unmatched checks) by
	\begin{equation} \label{eq:bar_f}
	{\bar J}_\text{S} = \sum_{m=1}^M \min\{0,\Delta_m\}.
	\end{equation}
	This energy function $\bar{J}_\text{S}$ can be simplified as $\dot{J}_\text{S}$ that counts the number of unmatched checks
	\begin{equation} \label{eq:dot_f}
	\dot{J}_\text{S} = \sum_{m=1}^M \min\{0,\sgn(\Delta_m)\} ~=\, -\sum_{m=1}^M |\hat z_m - z_m|,
	\end{equation}
	where $\hat z_m =\langle \hat E, S_m \rangle$.
	This energy function is used by a simplified BP called as \emph{bit-flipping BP} \cite{Gal63,SS96,Mac99,FMI99,RV20}, 
	which decides the update direction (which bit to flip) by minimizing the number of unmatched syndrome bits between $z$ and $\hat z$. 
	Bit-flipping BP has very low complexity and is useful for analyzing the convergence, since it only tracks the hard-decision information of the variable nodes.
	(Bit-flipping BP can also be used in practice, e.g., it was used in decoding expander codes \cite{SS96}.)

	To see the difference of the energy functions discussed above, we provide examples in Fig.~\ref{fig:Ex1_f}.

	For decoding the error pattern in Fig.~\ref{fig:surf_d7}\,(a), we plot the change of the energy function for the six configurations in Table~\ref{tbl:d7_degen1}. 
First, we consider the approximation $\tilde{J}_\text{S}$ in \eq{eq:tilde_f} with $B=6$, and the results are plotted in Fig.~\ref{fig:Ex1_f}\,(a). 
Conventional BP$_4$ (no matter parallel or serial) has achieved low energy, though the decoding is not successful.
If a larger step-size is used in normalized BP$_4$ (with $\alpha_c = 0.65$), the decoding then jumps randomly and diverges, resulting in high energy.
When using MBP$_4$ with $\alpha = 0.65$, the decoding converges to lowest energy. In this example, serial MBP$_4$ successfully converges to a degenerate error at iteration 10. 

Using \eq{eq:f_taylor_1} or \eq{eq:bar_f} results in a figure similar to Fig.~\ref{fig:Ex1_f_B}.
Thus we consider the energy function $\dot{J}_\text{S}$ in \eq{eq:dot_f}, and the results are plotted in Fig.~\ref{fig:Ex1_f}\,(b).
Obviously, only serial MBP$_4$ finally converges without getting trapped in local minima.

Note that, in Table~\ref{tbl:d7_degen1}, parallel BP$_4$ has a hard-decision pattern not changed after iteration 5. 
But the energy $\tilde{J}_\text{S}$ or $\dot{J}_\text{S}$ can still change, as shown in Fig.~\ref{fig:Ex1_f}. 
For example, consider the output distribution $(q_n^I,q_n^X,q_n^Y,q_n^Z)$ to oscillate between two points $(0.4, 0.3, 0.25, 0.05)$ and $(0.4, 0.05, 0.25, 0.3)$; then the hard-decision is the same $\hat E_n=I$ but, e.g., for an edge type $X$, the probability $q_n^I+q_n^X = 0.7 > 0.5$ for the first point and $q_n^I+q_n^X = 0.45 < 0.5$ for the second point, which can cause different energy levels.

There are two further notes.
First, although parallel MBP$_4$ gets trapped, it achieves low energy in both figures.
This explains why BP with post-processing usually works.

Second, when we try to plot $J = J_\text{D} + \eta J_\text{S}$ in \eq{eq:energy} rather than $J_\text{S}$ in \eq{eq:energy2}, it needs a quite large $\eta\approx 10^6$ to have a correct trend;
or otherwise, a successful decoding may result in high energy level.
This matches the expectation as follows. 
When decoding a highly-degenerate code, $\eta$ should be large to focus more on $J_\text{S}$ (since any degenerate errors can lower it) 
rather than $J_\text{D}$ (since a single low-weight error can dominate it, even if the error has incorrect syndrome).

\section{Color codes} \label{sec:color}
	Compared to the results for surface and toric codes (Table~\ref{tbl:thrd}), the results for color codes are as follows.
	The decoding problem of color codes can be cast as a hypergraph matching problem and approximately solved by MWPM  with a threshold of 13.3\% over depolarizing errors \cite{WFHH10}.
	In addition, a color code can be projected onto two surface codes and decoded by RG-BP with a threshold of 8.7\% over bit-flip errors \cite{BDP12}.
	A color code can be also projected onto three surface codes and decoded by MWPM with a threshold of 8.7\% \cite{Del14}, 
	or 8.4\% if decoded by UF \cite{DN17}, both over bit-flip errors.
	%
	Alternatively, without the need of the projection, color codes can be decoded by RG-BP, with a threshold of 7.8\% over bit-flip errors \cite{SR12}.
	Theoretical estimation suggests that a color code family can have a threshold of roughly 10.9\% over bit-flip errors \cite{Ohz09,KBM09}.
	For more information on the thresholds of various decoders, see \cite{LAR11,Ste14a}.

	For reference, AMBP$_4$ on color codes has a threshold of roughly 14.5\% over depolarizing errors \cite{KL22isit}.

\section{Linear-domain MBP} \label{sec:MBP_LD}

Algorithm~\ref{alg:QBP_M} provides the \ourBP in linear domain. 
The practical complexity can be improved as in Remark~\ref{rmk:cmplx}.


	\begin{algorithm}
		\begin{flushleft}
			\caption{: MBP$_4$ in linear domain} \label{alg:QBP_M}
			{\bf Input}: 
			$S \in\{I,X,Y,Z\}^{M\times N}$, $z \in\{0,1\}^M$, $T_{\max}\in \mathbb Z_+$, an $\alpha > 0$, 
			and initial probabilities $\{(p_n^I, p_n^X, p_n^Y, p_n^Z) \in \mathbb R^4\}_{n=1}^N$.

			{\bf Initialization.}  
			For $n\in\{1,2,\dots,N\}$ and $m\in\sM(n)$, let 
			\begin{equation*}
			d_{n\to m}=q_{n\to m}^{(0)}-q_{n\to m}^{(1)}, 
			\end{equation*}
			\begin{itemize} 
				\item[] where $q_{n\to m}^{(0)} = p_{n}^I+p_{n}^{S_{mn}}$ and $q_{n\to m}^{(1)} = 1 - q_{n\to m}^{(0)}$. 
			\end{itemize}
			
			{\bf Horizontal Step.} For $m\in\{1,2,\dots,M\}$ and $n\in\sN(m)$,\\
			~~~~~~ compute
			\begin{equation*}
			\delta_{m\to n} = (-1)^{z_m}\prod_{n'\in\sN(m)\setminus n} d_{n'\to m}.   
			\end{equation*}
			
			{\bf Vertical Step.} For $n\in\{1,2,\dots,N\}$ and $m\in\sM(n)$, do: 
			\begin{itemize}
				\item Compute 
				\begin{align}
				r_{m\to n}^{(0)} &= (\tfrac{1+\delta_{m\to n}}{2})^{1/\alpha}, ~~ r_{m\to n}^{(1)} = (\tfrac{1-\delta_{m\to n}}{2})^{1/\alpha}, \label{eq:rmn}\\
				q_{n\to m}^W &= p_n^W\prod_{m'\in\sM(n)\setminus m} r_{m'\to n}^{(\langle W, S_{m'n}\rangle)}, ~ W\in\{I,X,Y,Z\}. \notag 
				\end{align}
				\item Let
				\begin{equation} \label{eq:qnm}
				\begin{aligned}
				q_{n\to m}^{(0)} &= a_{mn}\,(q_{n\to m}^I + q_{n\to m}^{S_{mn}}) \,/\, (\tfrac{1+\delta_{m\to n}}{2})^{1 - 1/\alpha},\\
				q_{n\to m}^{(1)} &= a_{mn}\,(\textstyle \sum_{W'}q_{n\to m}^{W'}) \,/\, (\tfrac{1-\delta_{m\to n}}{2})^{1 - 1/\alpha},
				\end{aligned}
				\end{equation}
				where $W'\in\{X,Y,Z\}\setminus S_{mn}$ and $a_{mn}$ is a chosen scalar such that \mbox{$q_{n\to m}^{(0)}+q_{n\to m}^{(1)}=1$}.
				\item Update: $d_{n\to m} = q_{n\to m}^{(0)} - q_{n\to m}^{(1)}$. 
			\end{itemize}
			
			{\bf Hard Decision.} For $n\in\{1,2,\dots,N\}$, compute
			\begin{align*}
			~~~q_n^W = p_n^W \prod_{m\in\sM(n)} r_{mn}^{(\langle W,S_{mn}\rangle)}, ~ W\in\{I,X,Y,Z\}. 
			\end{align*}
			\begin{itemize}
				\item[]	Let $\hat E = \hat E_1\hat E_2\cdots\hat E_N$, where $\hat E_n = \argmax\limits_{W\in\{I,X,Y,Z\}} q_n^{W}$.
			\end{itemize}
			
			\begin{itemize}
				\item If $\langle \hat E, S_m \rangle = z_m ~\forall~ m$, halt   and return ``CONVERGE'';
				\item otherwise, if the maximum number of iterations $T_{\max}$ is reached, halt   and return ``FAIL'';
				\item otherwise, repeat from the horizontal step.
			\end{itemize}

		\end{flushleft}
	\end{algorithm}

\end{document}